\documentclass[english,structabstract]{aa}
\usepackage{mathptmx}
\usepackage[T1]{fontenc}
\usepackage[varg]{txfonts}
\usepackage{babel}
\usepackage{array}
\usepackage{textcomp}
\usepackage{multirow}
\usepackage{amsmath}
\usepackage{amssymb}
\usepackage{amsfonts}

\usepackage{graphicx}
\usepackage{natbib}
\bibpunct{(}{)}{;}{a}{}{,} %% natbib format for A&A and ApJ

\newcommand{\lyxmathsym}[1]{\ifmmode\begingroup\def\b@ld{bold}
  \text{\ifx\math@version\b@ld\bfseries\fi#1}\endgroup\else#1\fi}

\bibpunct{(}{)}{;}{a}{}{,} % to follow the A&A style
\makeatother

\begin{document}
%    \makeatletter \@removefromreset{figure}{chapter}\makeatother
    
\title{Constraining the thick disc formation scenario of the Milky Way}

\subtitle{}

\author{A.C. Robin\inst{1} \and  C. Reyl\'e\inst{1} \and J. Fliri \inst{2,3}  \and M. Czekaj\inst{4} \and C.P. Robert \inst{5}   \and A. M. M. Martins \inst{1} 
}

\institute{Institut Utinam, CNRS UMR6213, Universit\'e de Franche-Comt\'e, OSU THETA Franche-Comt\'e-Bourgogne, Observatoire de Besan\c{c}on, BP 1615, 25010 Besan\c{c}on Cedex, France  \\\email{annie.robin@obs-besancon.fr}
\and
Instituto de Astrof\'isica de Canarias, E-38200 La Laguna, Tenerife, Spain
\and
Universidad de La Laguna, Dept. Astrof\'isica, E-38206 La Laguna, Tenerife, Spain
\and
Departament d'Astronomia i Meteorologia and IEEC-ICC-UB,
     Universitat de Barcelona,
     Mart\'i i Franqu\`es, 1, E-08028 Barcelona, Spain 
\and
Universit\'e Paris-Dauphine, CEREMADE, 75775 Paris cedex 16, France
}

\offprints{A.C. Robin}

\date{Received ...; Accepted...}

\date{}

\abstract{}{More than 30 years after its discovery, the thick disc of the Milky Way is not fully explored. We examine the shape of the thick disc in order to gain insight into the process of its formation.} 
{The shape of the thick disc is studied in detail using photometric data at high and intermediate latitudes from SDSS and 2MASS surveys. We adopted the population synthesis approach  using an ABC-MCMC method to determine the potential degeneracies in the parameters, that can be caused by the mixing with the halo and with the thin disc. {We characterised the thick disc shape, scale height, scale length, local density, and flare, and we investigated the extent of the thick disc formation period by simulating several formation episodes. }}
{{We find that the vertical variation in density is not exponential, but much closer to a hyperbolic secant squared. Assuming a single formation epoch, the thick disc is better fitted with a ${\mathrm sech}^2$ scale height of 470 pc  and a scale length of 2.3 kpc. 
However, if one simulates two successive formation episodes, which mimicks an extended formation period, the older episode has a higher scale height and a longer scale length than the younger episode, which indicates a contraction during the collapse phase. The scale height ranges from 800 pc to 340 pc, the scale length from 3.2 kpc to 2 kpc.
The likelihood is much higher when the thick disc formation extends over a longer period. We also show that star formation increases from the old episode to the young and that there is no flare in the outskirt of the thick disc during the main episode. We compare our results with formation scenarios of the thick disc.}
During the fitting process, the halo parameters are determined as well. If a power-law density is assumed, it has an exponent of 3.3 and an axis ratio of 0.7. Alternatively, a Hernquist shape would have an exponent of 2.76, an axis ratio of 0.77, and a core radius of 2.1 kpc. The constraint on the halo shows that a transition between an inner and outer halo, if it exists, cannot be at a distance shorter than about 30 kpc, which is the limit of our investigation using turnoff halo stars. Finally, we show that extrapolating the thick disc towards the bulge region {explains well the stellar populations observed there, that there is no longer} need to invoke a classical bulge.}
{{The facts that the thick-disc episode lasted for several billion years, that a contraction is observed during the collapse phase, and that the main thick disc has a constant scale height with no flare argue against the formation of the thick disc through radial migration. The most probable scenario for the thick disc is that it formed while the Galaxy was gravitationally collapsing from well-mixed gas-rich giant clumps that were sustained by high turbulence, which prevented a thin disc from forming for a time, as proposed previously. This scenario explains well the observations in the thick-disc region and in the bulge region.}}


\maketitle

{}
%________________________________________________________________

\newcommand{\Msun}{$M_\odot~$}
\renewcommand{\deg}{$^{\circ}$}
\newcommand{\Ro}{$R_{\odot}~$}
\newcommand{\Rgal}{$R_{\mathrm{gal}}~$}

\section{Introduction}

Despite the fact that an intermediate population between the thin disc and halo has been identified more than 30 years ago \citep{Gilmore1983}, the so-called thick-disc properties are not yet agreed upon. Because of its intermediate characteristics, typically its density, scale height,  metallicity, and age, the thick disc is difficult to separate from the thin disc on the one hand, which dominates the stellar density close to the Galactic plane, and from the halo on the other hand, which dominates the star counts at large distances. The thick disc has been investigated by trying to distinguish its members by their kinematics, by their position (at intermediate distance from the plane), and by its abundances. These analyses have led to confusion because the selections do not recover a clearly identified unique population. They all suffer from either selections that bias the determined mean and dispersion, or from contaminations from the thin disc or the halo. 

Is there a distinct thick disc at all ? This question has found different answers over the years. \cite{1984ApJS...55...67B} claimed that such a population is not needed; at the same time \cite{Gilmore1983} and \cite{Robin1986} answered this question with a convincing 'yes'. To concentrate on more recent evidence, \cite{Veltz2008} analysed RAVE data to show that part of the thin disc has the characteristics of a thick disc, and is clearly separated in kinematics. Most recently, the question was raised again when \cite{Bovy2012b} claimed that there is no distinct thick disc, but instead a continuity between the traditional thin disc and a population with a thicker density distribution, higher velocity dispersions, and lower metallicities. One argument for this finding was the continuity in the [$\alpha$/Fe] vs [Fe/H] relation in the SEGUE sample. However, more recent data, in particular the sample assembled by \cite{Adibekyan2013} in the solar neighbourhood, clearly showed two sequences across the metallicity range [-1; 0], one with high [$\alpha$/Fe] values corresponding to the thick disc, and one with solar  [$\alpha$/Fe] corresponding to the thin disc, as claimed by \cite{Haywood2013}. This means that the alpha element abundances apparently are the key to separating the thick disc from the thin disc. These abundances currently seem to be the best proxy for ages in field populations. The position of a population in the   [$\alpha$/Fe] vs [Fe/H]  is directly linked with the history of star formation. 

The question however is more to discover how these populations form and interact than to discuss whether the disc should be separated into 2, 5, or 10 sub-components.
One can argue that the thick disc is just a part of the disc, or that it traces an epoch of star formation that differs in intensity from the main thin disc, or was separated by an episode of no or low star formation. It might also originate from a totally different physical process. Several scenarios have been proposed over the years. Today the most popular ones are i) relics of early mergers; ii)  thin-disc thickening by mergers; 
iii) thin-disc thickening by bar and/or spiral instabilities; { iv) radial migration; and v) giant turbulent gas clumps at high z}.

To be able to solve the thick disc, and whole disc, origin, a better understanding of the mass distribution and the way these populations are distributed as a function of age are key points one needs to solve to distinguish between the scenarios. 
Another crucial element in understanding the disc formation today is the chemical enrichment derived from
abundance ratios, alpha elements over iron in particular, and their distribution and variations in the Galaxy. But the analysis is made more complex because in most of the samples the birth places of the stars are located in different regions, and radial and vertical migrations need to be taken into account. The importance of migrations has been shown by \cite{Sellwood2002} and has been studied in different N-body simulations (\cite{Roskar2012, Minchev2011,Dimatteo2011,Bird2012} among others). These simulations do not yet create a unique picture. But they help identifying the different processes that can contribute to the thick-disc formation as well as the signatures that are expected in available surveys.

Spectroscopic data are most important for analysing the characteristics of the populations in different locations of the galaxy in detail and in situ. However, they do not yet provide a direct unbiased measurement of the distances, which still depend on stellar models. Moreover, they generally suffer from severe selection functions, which are very difficult to correct for. Therefore it is more difficult to achieve the mass distribution of a given population in incomplete spectroscopic surveys than from complete photometric surveys.

With this point of view in mind, we here attempt to analyse the spatial distribution of the populations at medium and high latitudes in the Galaxy to be able to characterise their mass density. Because we use the population synthesis approach, we assume the age and metallicity distributions of these populations, that, at the end, are constrained as well, but not as accurately as with spectroscopic surveys.

We aim at constraining the characteristics of the thick-disc population of the Milky Way: space distribution, age, and mean metallicity using stellar statistics of the Sloan Digital Sky Survey (SDSS) data release 8  \citep{SDSSDR8} and the Two Micron All Sky Survey (2MASS)  \citep{Skrutskie06}. This approach is complementary to analyses that study the correlations between various tracers of Galactic evolution, such as the kinematics and metallicity, as well as the stage of evolution of the stars. The density distributions of the populations can help in understanding the true shape and density of each population and to compute the Galactic mass distribution and gravitational potential. The densities are difficult to reconstruct from spectroscopic samples because of the way the samples are built. Using photometrically complete samples helps to better understand the underlying distribution. { Derived distances of the stars are often a source of bias and uncertainties because of the difficulty in distance estimations from photometry. Even in spectroscopy the errors on the gravity are often large enough to induce biases on the result. Instead, with the population synthesis approach, we use the stellar density and colour distributions in many directions to deduce the distance scales of various populations}. Finally, we deal with much larger samples than are presently available from spectroscopic surveys. 

In Sect.~2 we present the selection of data, ensuring that the sample is not contaminated by any significant stream. Sect.~3 is devoted to the model presentation and the main parameters and assumptions. Sect.~4 presents the fitting method based on a Markov chain process, while Sect.~5 shows the resulting fits and the sensitivity to input parameters. We discuss the effect of the halo shape and thin-disc characteristics on the results in Sect. 6. { We present a way to simulate a longer star formation period for the thick disc in Sect. 7 and show how it improves the fit. We explore the consequences for the populations in the inner bulge in Sect. 8. Sect.9 discusses our results and conclusions are given in Sect. 10.}

\section{Data selection}

\begin{figure*}[htb]
\begin{center}
\includegraphics[width=12cm,angle=0,trim=0 150 0 150,clip=true]{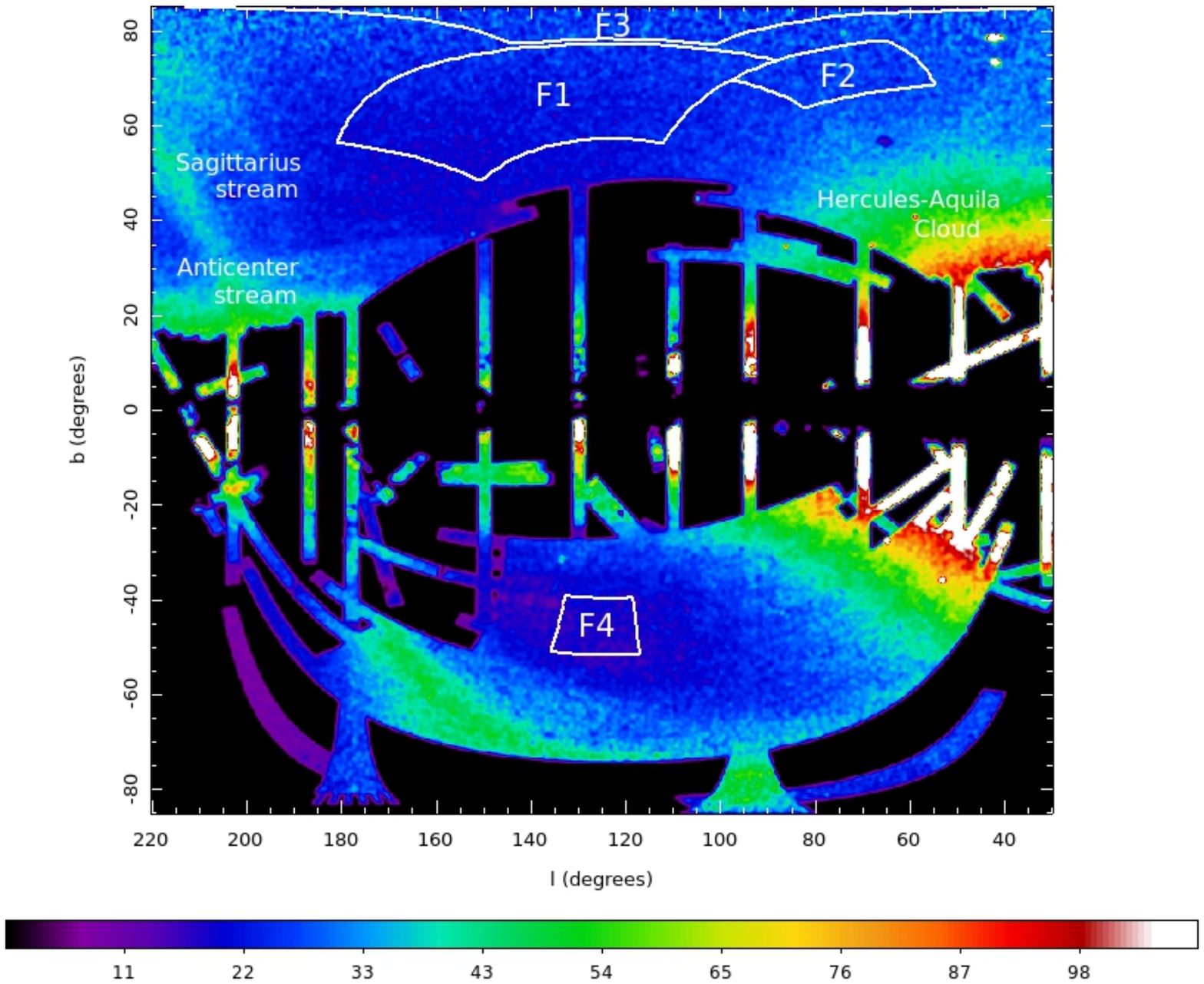}
\caption{SDSS DR8 density map of main-sequence turnoff stars at a resolution
of 0.25 degrees, showing the outline of fields F1 to F4 in Galactic
coordinates. The selected fields avoid significant stellar streams and
overdensities such as the Sagittarius stream, the Anticentre/Monoceros
structure or the Hercules-Aquila cloud.}
\label{fig_SDSS}

\end{center}
\end{figure*}

Photometric data from the SDSS-DR8 were considered first.
To fit the smooth part of the Galaxy we tried to avoid
regions where significant streams are known. Based on stellar density
maps, four patches were selected, which sample a considerable range in
Galactic latitude between $b_{min}=48$\deg~ and $b_{max}=90$\deg~ for the
northern fields (F1 -F3) and $b_{min}=-52$\deg~ and $b_{max}=-39$\deg~ for field
F4 in the southern hemisphere. The areas are given in Table~\ref{SDSS_fields} 
and are shown in Fig.~\ref{fig_SDSS}. 
The fields are considered to be free of
significant streams (Sagittarius, Orphan, and Anticentre streams) and
stellar overdensities (Virgo overdensity and the Hercules-Aquila
cloud) up to heliocentric distances of approximately 40 kpc. DR8 data
were obtained from the SDSS archive using the SQL based query system
{\it CasJobs}, by selecting stars with clean photometry in $g$, $r$ and $i$
and by using a combination of photometry flags\footnote{see 
\url{http://sdss.lib.uchicago.edu/dr7/en/help/docs/realquery.asp}}.

We selected stars with $15<r<21$ and applied colour cuts to avoid contamination by compact galaxies and quasars. The final selection is

\noindent $-0.25<r-i-0.45*(g-r)<0.15 $ \\
\noindent $u-g+1.2*(g-r) >0.8$ \\
\noindent $(g-r)<1.2$. \\

The selection in g-r colour lower than 1.2 avoids contamination by thin-disc stars. It also avoids the part of the HR diagram where the stellar models become more uncertain.  
The selection covers a large part of the halo and thick-disc colour-magnitude diagrams (CMD), significantly lower than the turnoff, down to about M$_{V}$=8.

Each patch was separated into smaller sub-fields where the star counts were performed.  Each individual field covers 5 $\times$ 5 square degrees for patch F1 and 4 $\times$ 4 square degrees for the three others.
In each sub-field star counts were performed in the colour magnitude diagram space (r, g-r) to be able to constrain the density of populations in different parts of the HR diagram, which essentially is a function of the luminosity function and isochrone of the population on the one hand and of the density law as a function of position in the Galaxy on the other hand.

\begin{table}
\caption{Sky coverage of the four selected SDSS patches and the number of sub-fields considered for producing the star counts. Sub-fields cover 5\deg$\times$5\deg~ in patch F1 and 4\deg$\times$4\deg~ in patches F2 to F4.}
\label{SDSS_fields}
\begin{tabular}{lccc}
\hline
 & RA range & DEC range &Sub-fields \\
\hline
F1 &  155\deg - 205\deg & 40\deg - 60\deg   &10 $\times$ 4\\
F2 & 205\deg - 217\deg & 33\deg - 45\deg  &3 $\times$ 3 \\
F3 & 187\deg - 199\deg &  27\deg - 39\deg & 3 $\times$ 3 \\
F4 & 9\deg - 21\deg & 11\deg - 23\deg &  3 $\times$ 3 \\
\hline
\end{tabular}
\end{table}

\begin{figure}[htb]
\begin{center}
\includegraphics[width=9cm,angle=0]{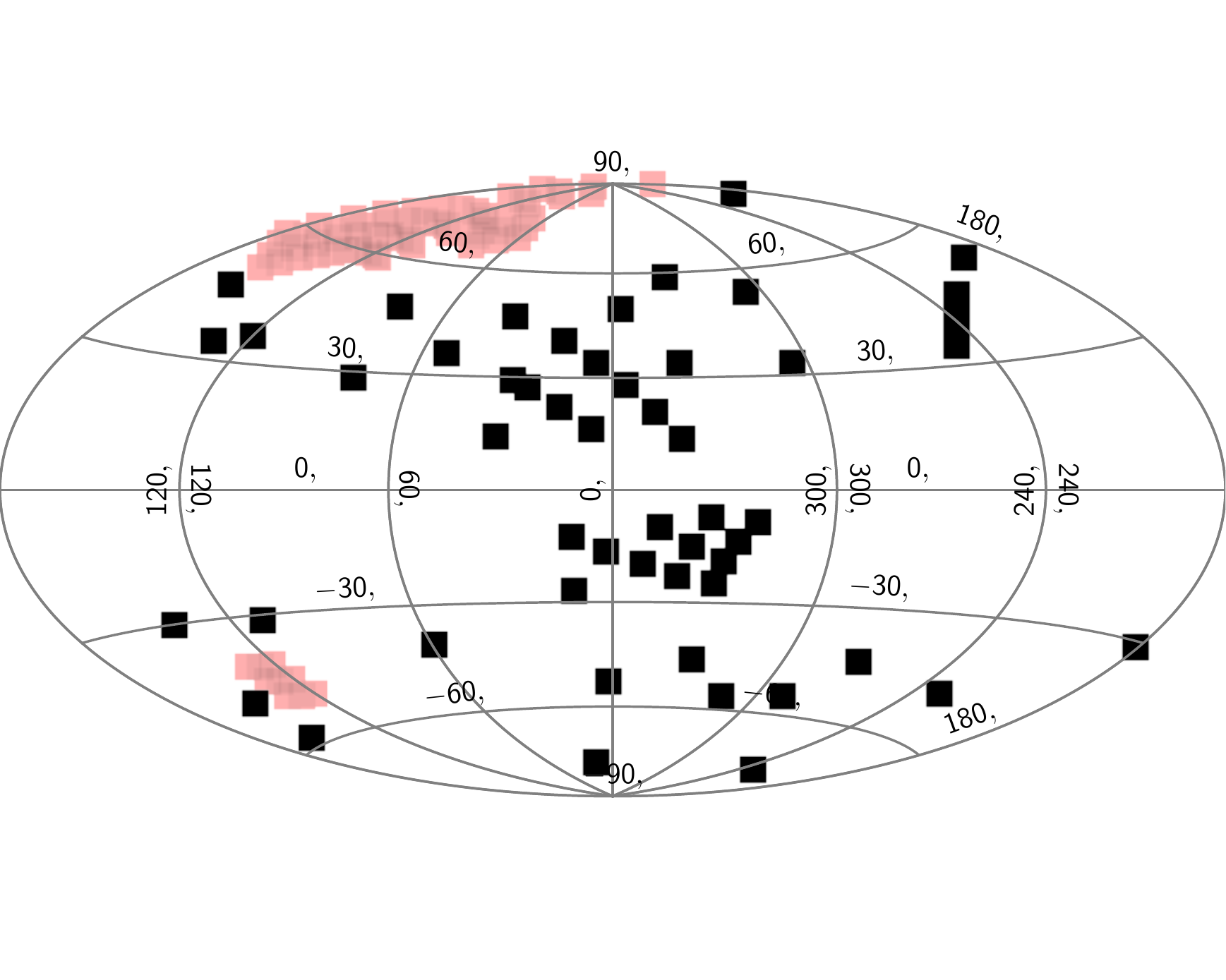}

\caption{Location of the 54 fields from the 2MASS survey (black) and 67 sub-fields from the SDSS (grey-pink) in Galactic coordinates.}
\label{fields}
\end{center}
\end{figure}

The first attempts to fit the model parameters using SDSS fields alone showed that the fit was not sensitive enough to constrain the thick-disc scale length. This result was also obtained by \cite{Juric2008}. This effect is stronger in our case because we eliminated regions contaminated by streams. 
Accordingly we decided to add 2MASS fields at intermediate latitudes and at longitudes covering the inner galaxy, and also closer to the plane in the outer Galaxy. Fig.~\ref{fields} presents the distribution on the sky of the fields (SDSS and 2MASS) used for the model fitting. The 2MASS fields are much shallower and do not help much in constraining the halo density, but they give a strong constraint on the thick disc, as seen below, and are less contaminated by streams.

Fig.~\ref{Rz-plane} gives an overview of the distribution of the simulated stars in the Galactocentric cylindrical coordinates (R,z).

\begin{figure}[htb]
\begin{center}
\includegraphics[width=9cm,angle=0]{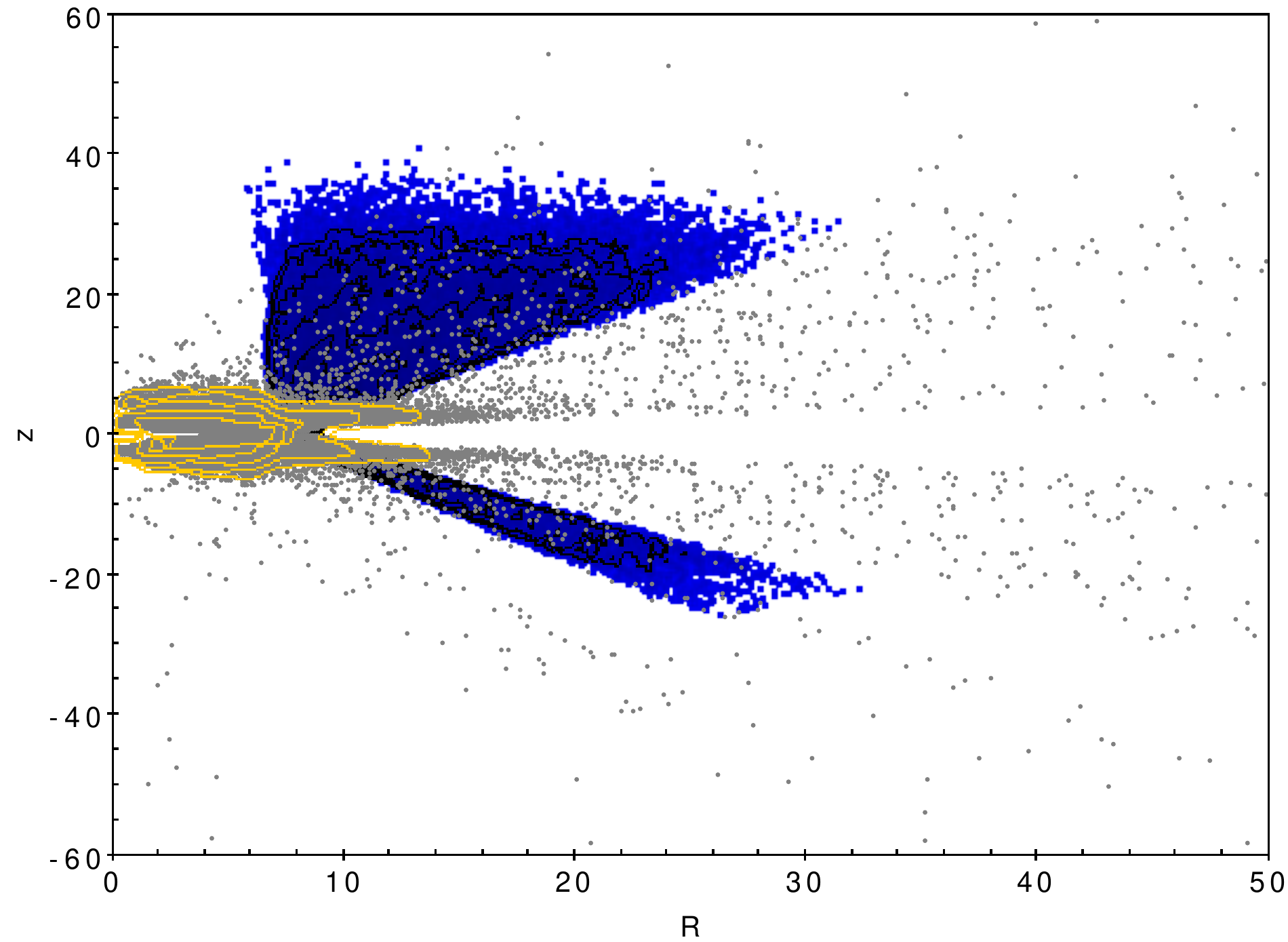}
\caption{Coverage of the (R,z) Galactocentric cylindrical coordinate plane of the SDSS (blue dots and black contours) and 2MASS fields (grey dots and yellow contours). For the sake of producing this plot from SDSS fields we selected subgiants near the turnoff, and for 2MASS fields, K and M giants. In SDSS field giants are located farther away but with lower densities.}
\label{Rz-plane}

\end{center}
\end{figure}

\section{Basic model ingredients}

The Besan\c{c}on Galaxy Model \citep{Robin2003} (hereafter BGM) was used to provide the basic simulations, which were modified to change the shapes of the input populations. This model  has undergone a few changes recently \citep{Robin2012a}. We list here the changes that can play a role in the fit.

\subsection{Megacam to SDSS photometric system}

In \cite{Schultheis2006} we presented the calibration of the model in the Megacam photometric system, which is valid for the observations with the Megacam mosaic at CFHT and applies to the survey CFHT-LS\footnote{\url{http://www.cfht.hawaii.edu/Science/CFHTLS/}}. 

To compute the SDSS photometry, we used colour equations for transforming the photometry from Megacam system to SDSS because the two systems are very similar, specially for filters g, r, and i. The colour equations were taken from the Supernova Legacy Survey \citep{Regnault2009}.  There can be slight shifts of zero points between Megacam system and Sloan photometry, but these zero points vary from field to field, as noted in the CFHT-LS web site\footnote{\url{http://terapix.iap.fr/cplt/T0007/doc/T0007-doc.html}}. The shifts are generally smaller than 0.02 magnitudes in g,r,i and z bands, but can amount to -0.02 to -0.05 in the u band. We did not apply these shifts in the simulation of SDSS data, because we did not use the u band { for the comparison between model and data, but only to clean the SDSS sample from contamination by extragalactic sources}. An attempt to use a better calibration from model atmospheres, following the work by \cite{An2009} was made, but the model grid they use does not cover the whole range of effective temperatures, gravities, and metallicities that are present in our simulations and cannot be extrapolated further. For example, giants are not considered but are important in the present analysis. Moreover, the metallicity range is limited to [Fe/H] $>$ -2 dex, which is a problem for the halo. We checked that for the colour used in our analysis (g-r) a simple colour equation from \cite{Regnault2009} is sufficient and gives similar results to those of \cite{An2009} in their temperature range.

The transformations used in the present study are given in  Eqn.~\eqref{mega2sdss}.
\begin{eqnarray}
 u_{\mathrm SDSS} - u_{\mathrm M} &=& 0.2674 \times (u_{\mathrm M} - g_{\mathrm M}) 
           -0.049 \times  (g_{\mathrm M} - r_{\mathrm M})  \nonumber \\
 g_{\mathrm SDSS} - g_{\mathrm M} &=& 0.1833 \times (g_{\mathrm M} - r_{\mathrm M})
             -0.0051\times   (r_{\mathrm M} - i_{\mathrm M}) 	        \nonumber \\  
 r_{\mathrm SDSS} - r_{\mathrm M} &=& -0.0280 \times (r_{\mathrm M} - i_{\mathrm M}) \nonumber \\
 i_{\mathrm SDSS} - i_{\mathrm M} &=& 0.0951  \times (r_{\mathrm M} - i_{\mathrm M}) \nonumber \\
 z_{\mathrm SDSS} - z_{\mathrm M} &=& -0.0373 \times (i_{\mathrm M} - z_{\mathrm M})
        - 0.0035\times  (r_{\mathrm M} - i_{\mathrm M})    
  \label{mega2sdss}
 \end{eqnarray}

where the indice $M$ stands for Megacam filters.

\subsection{Thin-disc local density and dynamical self-consistency}

The thin disc does not play a crucial role in the present analysis, because of the colour cut (this is demonstrated in the discussion in sect.~6). However, it is useful to evaluate the contribution of the thin disc and make it as realistic as possible. Since publishing our previous analysis \citep{Robin2003} we reconsidered the disc local density and the dynamical self-consistency, which allows constraining the thin-disc density laws (particularly the density gradient perpendicular to the Galactic plane) by using the global Galaxy potential and assuming an age-velocity dispersion relation. The local mass density was recently revised by \cite{vanLeeuwen2007} by using a re-reduction of the Hipparcos catalogue. In contrast to \cite{Creze1998}, who found a local mass density of 0.076 \Msun pc$^{-3}$, \cite{vanLeeuwen2007} found a much higher value of 0.112 \Msun pc$^{-3}$. Today intermediate values given by \cite{Holmberg2004} of 0.102 \Msun pc$^{-3}$ and by \cite{Korchagin2003} of 0.100 \Msun pc$^{-3}$ are considered as the reference value. To match this value, we corrected the interstellar matter mass density and attributed to it 0.05 \Msun pc$^{-3}$, in agreement with \cite{Binney2008} and still within the error bars given by observational constraints. 
In the new computation of the dynamical consistency, the local mass density is summarized as follows: 0.041 \Msun pc$^{-3}$ for the stellar density, 0.05 \Msun pc$^{-3}$ for the interstellar matter, and 0.011 \Msun pc$^{-3}$ for the dark matter halo. The eccentricities of the thin disc are then slightly lower than before. They extend from 0.014 for the thinnest younger component to  0.070 for the thicker old disc.

\subsection {Thick-disc shape}

In \cite{Reyle2001} we showed that it was difficult to constrain the thick-disc density law well because of a degeneracy between its local density and scale height. Moreover, an exponential shape is often assumed for the thick disc, but is probably not realistic, even though it is a simple and easy way to simulate it. Here we
attempt to better constrain the thick-disc shape. This means that we constrain the way its density decreases as a function of distance perpendicular to the Galactic plane, as well as the local density, and its radial scale length. We also attempt to check whether the thick disc can be better modeled with scale height varying with Galactocentric radius, which is only possible if wide ranges of longitude and latitude are covered. We also consider a multi-age thick disc.

We essentially used two shapes for the thick disc. The first one (shape A) is what we assumed in previous studies. This is a density that decomposes vertically into a parabolic shape at short distances from the plane followed by an exponential (see Eqn~\eqref{eqn-thdisc}). This shape assumes four parameters: the local density $\rho_{0}$ at the solar position, the scale height of the exponential part $h_{z}$, the position of the change $\xi$, and the horizontal scale length  $h_{R}$.

\begin{equation}
\label{eqn-thdisc}
\rho(R,z) =  \left\{\begin{array}{ll}
\rho_{0} \exp{(-\frac{R-R_{\small \sun}}{h_{R}})}\times(1-\frac{z^{2}/h_{z}}{\xi*(2.+\xi/h_{z})}) & 
\mbox{ if } z\leq \xi \\
\rho_{0} \exp{(-\frac{R-R_{\small \sun}}{h_{R}})}\times\exp({-\frac{|z-z_{\small \sun}|}{h_{z}})\times  \frac{2\exp(\xi/h_{z})}{(2+\xi/h_{z})}} &
\mbox{ if } z> \xi,\\
\end{array}
\right.
\end{equation}

The second shape (shape B)  is a simple hyperbolic secant squared, which has three parameters: local density, scale height  $h_{z}$, and scale length  $h_{R}$. 

\begin{equation}
\label{eqn-thdisc}
\rho(R,z) = \rho_{0} \cosh(\frac{|z-z_{\small \sun}|}{2 h_z})^{-2} \times \exp(-\frac{(R-R_{\small \sun})}{h_{R}}),
\end{equation}

We also considered a pure exponential, but because the results were much poorer than the others (and also just a particular case of shape A when $\xi$ tends towards 0), we disregarded it.

Depending on the formation scenario of the thick disc, the vertical scale height can be found to be flaring in the outer Galaxy. We explored this possibility by adopting a linear flare.  In this case the scale height changes linearly with Galactocentric radius, starting at a Galactocentric radius $R_{flare}$. The two parameters to fit are then $R_{flare}$ and $g_{flare}$, the slope of the variation of scale height (see Eq. \eqref{eqn-flare}).

\begin{equation}
h_{z} = \left\{\begin{array}{ll}
h_{z,0}\times g_{flare} \times (R_{g}-R_{flare})& 
\mbox{ if }  R_{g}>R_{flare}\\
 h_{z,0} &
\mbox{ if }  R_{g}<=R_{flare}, \\
\end{array}
\right.
\label{eqn-flare}
\end{equation}

Notice that we did not exclude a negative value for $g_{flare}$ a priori, which would lead to a smaller scale height in the outer Galaxy.

\subsection{Halo shape}

At high latitudes, there is a smooth transition between the thick disc and the halo in the colour-magnitude diagrams. The turnoff appears to be dominated by the thick disc up to $r\approx17-18$, while at fainter magnitudes the halo dominates. But the transition is smooth enough and its exact magnitude changes as a function of longitude and latitudes for geometrical reasons. In our analysis we considered the star counts up to $r=$21. Therefore we must ensure that the halo is well simulated and does not bias our conclusion on the thick disc. In the range of magnitude we considered the halo does not reach very far distances at the turnoff (about 40 kpc). Giants contribute to the counts in the redder part of the CMD, although they are not dominant in any part of the CMD in the magnitude range considered ($r>$15). 

We considered several halo shapes: simple power-law, double power-law \citep[to simulate a dual halo to test the hypothesis from][]{Carollo2010}), and finally a Hernquist halo. For power laws, the halo shape was assumed to follow two simple power-laws with a break in between. Eq.~\eqref{Eq-halo} gives the parametrized halo density law. The single power-law is just a particular case where the $R_{Break}$ is larger than 150 kpc.

\begin{equation}
\label{Eq-halo}
\rho(R,z) = \left\{\begin{array}{ll}
\rho_{0} \times (R^2+(\frac{z}{q_1})^{2})^{\frac{-n_1}{2}} & 
\mbox{ if } R\leq R_{Break}\\
K \times  (R^2+(\frac{z}{q_2})^{2})^{\frac{-n_2}{2}} &
\mbox{ if } R> R_{Break},\\
\end{array}
\right.
\end{equation}

where (R,z) are Galactocentric cylindrical coordinates, $q_{i}$ is the axis ratio, and $n_{i}$ is the exponent of the power law, with $i=1$ for the inner halo defined as R$\leq R_{Break}$ and $i=2$ for the outer halo. K is a constant to ensure the continuity at $R_{Break}$.

For a Hernquist halo, we considered Eq.~\eqref{Eq-halo-Hernquist},

\begin{equation}
\label{Eq-halo-Hernquist}
\rho(R,z) = 
\rho_{0} \times \frac {1}{R_{a}\times(R_{{core}}+R_{a})^{n} },
\end{equation}

where $R_{a}=\sqrt{x^2+\frac{y}{p}^2+\frac{z}{q}^2}$ 

(x,y,z) are Galactocentric cartesian coordinates, p and q are axis ratios. In a triaxial Hernquist halo (p different from 1) three angles determine the orientation of the ellipsoid.

\subsection{Mass function}

Different initial mass function (IMF) can be considered for the thick disc. However, the range of masses to which these data are sensitive is very narrow, between 0.5 and 0.9 \Msun for the thick disc and even narrower for the halo, because most of our stars are close to the turnoff. We incorporated the thick disc IMF in our fits and found an IMF slope varying between $\approx$-0.2 and $\approx$0.2, compared with the slope of 0.6 for the corresponding masses in the thin disc. However, since the mass range covered is narrow it is difficult to conclude that the IMF of the thick disc is different from that of the thin disc. We also argue that since the binarity is not taken into account, if the probability of binaries were different in the thick disc, it might imply an effective IMF slope different from the thin disc, not because it is really different, but because the binarity effect is different. The IMF slope is not a major parameter for the analysis we performed here. It can change the value of the local density normalisation of the thick disc only slightly. We plan to conduct a deeper analysis of the thick-disc IMF and binarity rate in a future paper.

\subsection{Age, metallicity and stellar models}

Age and metallicity are two major parameters that define the distribution of the stars in the CMD. {The CMDs were generated from isochrones with an age in the range 9 to 13 Gyr and a mean metallicity ranging between -0.8 and -0.5 dex  for the thick disc and -1.5 dex for the halo. To generate the stars the metallicity was computed from a Gaussian with a dispersion of  0.3 dex for the thick disc and 0.5 dex for the halo}. These dispersions are consistent with the value reported by \cite{Ivezic2008} value for the thick disc but higher than the value quoted there for the halo. However, the simulated colour histograms at $g>19$ in which the blue wing is dominated by the halo, is well reproduced by this hypothesis, better than by a lower metallicity dispersion of 0.3 dex.

The stellar models used so far are based on the alpha-enhanced isochrones of \cite{Bergbush}. We alternatively attempted to use Padova isochrones \citep{Bertelli2008}. In practice, these are very similar to those of \cite{Bergbush} for the age and metallicity studied, but Padova isochrones present significant problems for low-mass stars. For this reason they were discarded. Basti isochrones \citep{Pietrinferni2004} were used for generating the blue horizontal branch for halo stars because this stage is missing in the models of \cite{Bergbush}. In practice, the horizontal branch has no influence on the present study.

{ We explored a range of values for the age and metallicity for the thick disc.
The range of ages studied here for the thick disc covers 8 to 13 Gyr.  For the metallicity we attempted to fit the thick-disc iron abundance from $6\times 10^{-3}$ to $1\times 10^{-3}$. It covers the range of expected metallicities for the thick disc well, including a metal-poor thick disc. For the halo we only considered an age of 14 Gyr.}

\subsection{Extinction treatment}

We first considered the \cite{Marshall2006} 3D extinction map, but only one field can be treated this way because their map does not explore latitudes higher than 10\degr~ in absolute value. For higher latitude fields, we considered a smooth extinction, modelled by a double exponential disc of scale height 140 pc and scale length 4500 pc. The local normalisation was assumed to be 0.7 mag/kpc, following \cite{Robin1986}. Simulations were made with this extinction, but a dispersion around the mean value was added on a star-by-star basis, equivalent to 20\% of the mean extinction, to account for small-scale variations of the interstellar matter distribution. For particular fields closer to the plane, we checked the extinction by comparing the simulated CMDs with observed ones, and revised them appropriately. For example, the normalisation of the diffuse extinction distribution was revised to 1.3 mag/kpc in Av in patch F4. { We did consider the maps of \cite{Schlegel1998} because they only give the 2D extinction value integrated over lines of sight and might suffer from systematics, as shown by \cite{Dobashi2005}, and more recently by \cite{Schlafly2011}, and by \cite{Berry2012} using SDSS data.}

\section{Fitting method}

The parameter space was explored using a Monte Carlo Markov chain technique. Since the model is complex, the likelihood cannot be analytically computed. Instead, we used the ABC (approximate Bayesian computation) method \citep{marin:pudlo:robert:ryder:2011}. 
For each set of parameters a distance was computed between the simulated data (star counts in bins of magnitude and colour) and the observations. This distance uses the formula given by \cite{Bienayme1987} for a binomial statistics \citep{Kendall1973} that is given in eqn.~\ref{likelihood},

 \begin{equation}
 Lr= \sum_{i=1}^{N} q_i \times (1. -R_i + \mathrm{ln}(R_i)), 
  \label{likelihood}
\end{equation}
 where {$Lr$ is the reduced likelihood for a binomial statistics}, $i$ is the index of the bin,  $f_i$ and $q_i$ are the number of stars in bin $i$ in the model and the data, and $R_i=\frac{f_i}{q_i}$.
 
Then the exponential of minus this distance was used as a nonparametric estimator of the probability density and of the likelihood function.  
The ABC-MCMC algorithm was implemented using this non-parametric estimate in the Metropolis-Hastings algorithm acceptance ratio (Metropolis et al. 1953; Hastings 1970). The scale factor in the proposal distribution used in the Metropolis-Hastings algorithm was chosen in such a way that the overall frequency of acceptance stands between 5 and 25\%.

Models with a different number of parameters than the reference model (thick-disc shape A) were compared using the  Bayesian information criterion (BIC) \citep{Schwarz1978} that penalises larger models by
a factor $k\times ln(n)$, where k is the number of parameters and n the number of observations (Eqn.~\ref{BIC}),

 \begin{equation}
 BIC=-2.\times Lr + k\times ln(n).
  \label{BIC}
\end{equation}

In our case $n$=14830. For the reference model $k$=10, but varies between 10 and 17 in other models.

We made about 5 to 20 runs for each model tested, to verify the convergence. The two shapes of the thick disc were tested separately, as well as each age/metallicity combination.  The explored parameters and their respective ranges  are presented in Table~\ref{table_t0m12b-12g3} 
for the shape A thick disc with an age of 12 Gyr and a metallicity of z=0.003. 

The parameters determined for each run were then compared and checked for correlations. An estimation of the uncertainty on each parameter was based on the final batch of each Markov chain and on a comparison of parameters in independent runs.
Finally, CMD and histograms were plotted for the best likelihood model and checked for remaining systematics.

\section{Results}

We studied different age and metallicity combinations for the thick disc and obtained a maximum-likelihood estimate for each of them. By default, the halo was assumed to follow a single power-law. Alternatives are discussed below. In this case the thick disc with an age of 12 Gyr and a metallicity of z=0.003 gives the best result. We show first the parameters for the best likelihood, together with the acceptable range, for this best age/metallicity combination and for the two thick-disc shapes. Then we show how the likelihood and the fitted parameters vary when we change the age and metallicity of the thick disc. 

The implications of changing the halo shape or the thin disc are explored in the next section.

\subsection{Shape A thick disc of 12 Gyr and z=0.003}

Fitting results for a thick disc of shape A, age of 12 Gyr and z=0.003 are presented in Table~\ref{table_t0m12b-12g3}. In this test the halo was assumed to follow a single power-law. 
 
\begin{center}
\begin{table}
\caption{Parameters for a thick disc of shape A: fitted model parameters, range explored, best likelihood value, and estimate of the standard deviation for the thick disc of shape A, an age of 12 Gyr, and a metallicity z=0.003
 }
\label{table_t0m12b-12g3}

\begin{tabular}{lllll}
\hline
Parameter & Min  & Max  & Best & StDev\\
\hline
{\it Halo} \\
exponent & 2.2 & 4.5                & 3.36 & 0.02\\
axis ratio & 0.3 & 1.2             &     0.757 & 0.007\\
normalisation & 0. & 5. & 2.82 & 0.04 \\
\hline
{\it Thick disc }\\
 scale height (pc) & 450 & 1300           & 535.2 & 4.6\\ 
 normalisation & 0.  &  3. &  1.647 & 0.059 \\
IMF slope & -1 & 1 & -0.22 & 0.05\\
 scale length (pc) & 1500 & 5000          & 2362. & 25.\\
 $\xi$ (pc) & 100 & 1500                &      662.8 & 10.4\\
 flare start radius (pc) & 7000 & 20000 & 9261. & 324.\\
 flare slope (pc/kpc) & -0.3 & 0.3 & 0.1310 & 0.069\\
  Lr & & & -64485. & 484.\\
  BIC & & & 129056.4 & \\
  \hline
\end{tabular}

\end{table}
\end{center}

\begin{figure}

\begin{tabular}{c}
   \includegraphics[width=6cm]{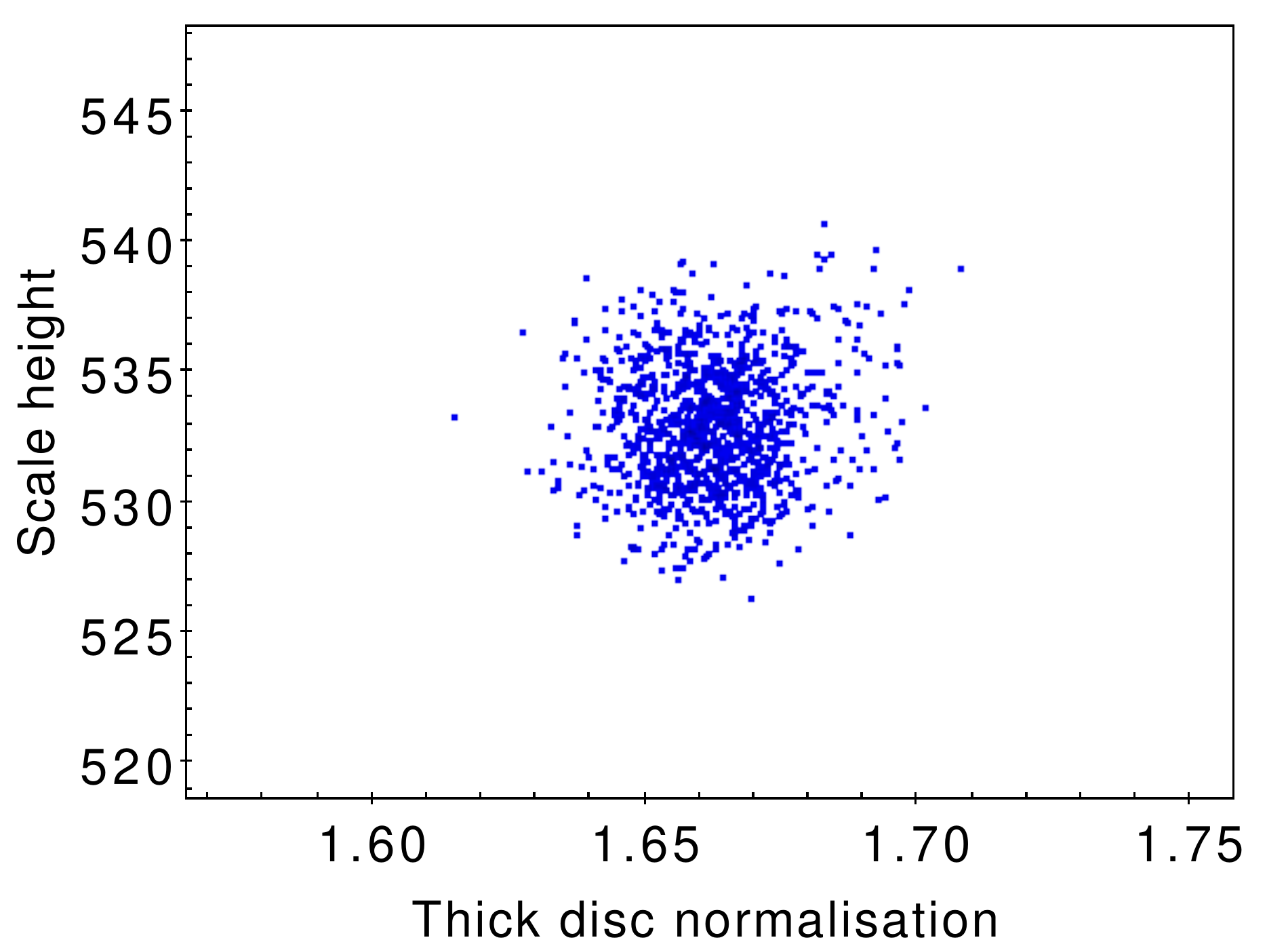} \\
   \includegraphics[width=6cm]{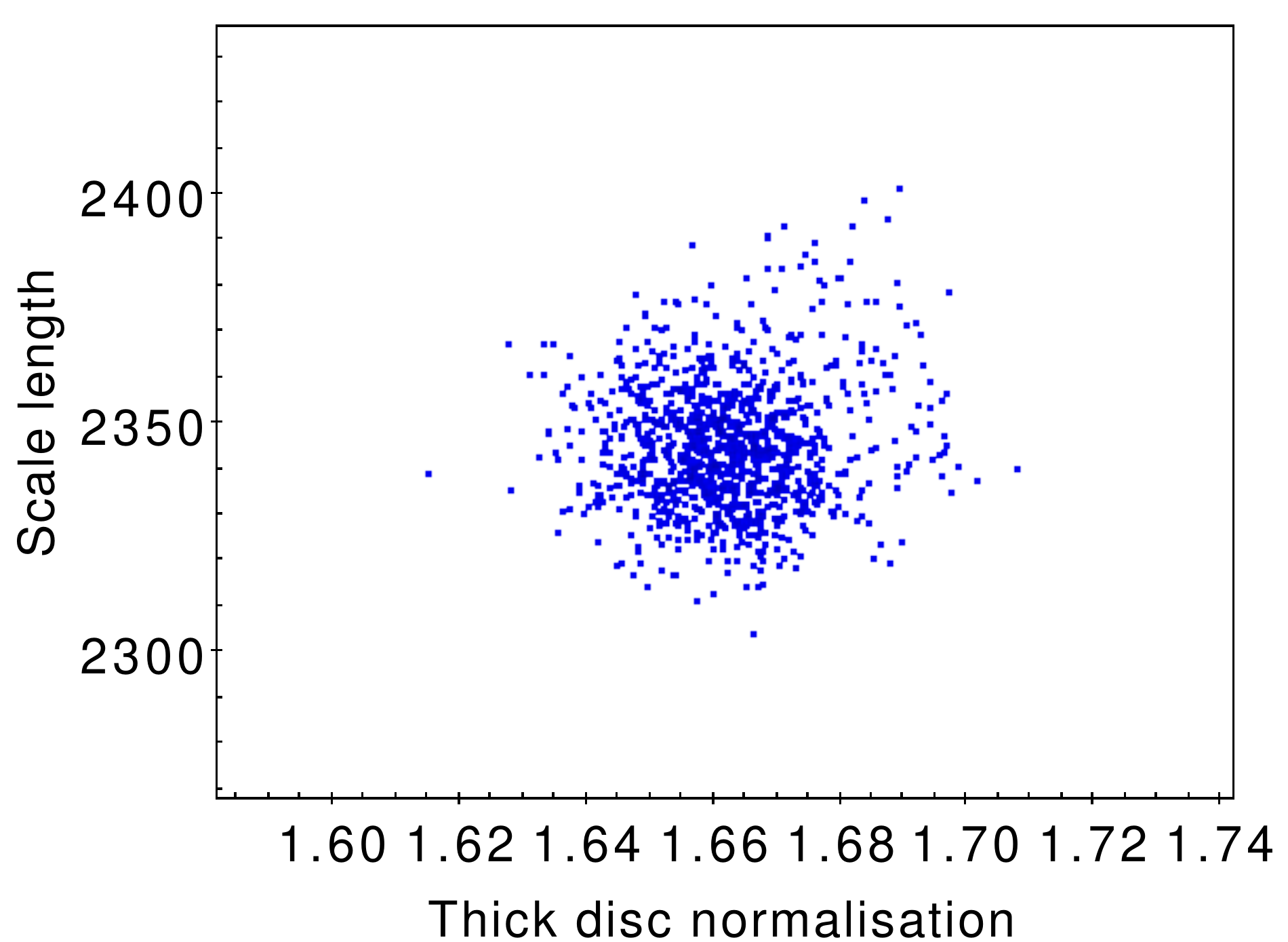} \\
    \includegraphics[width=6cm]{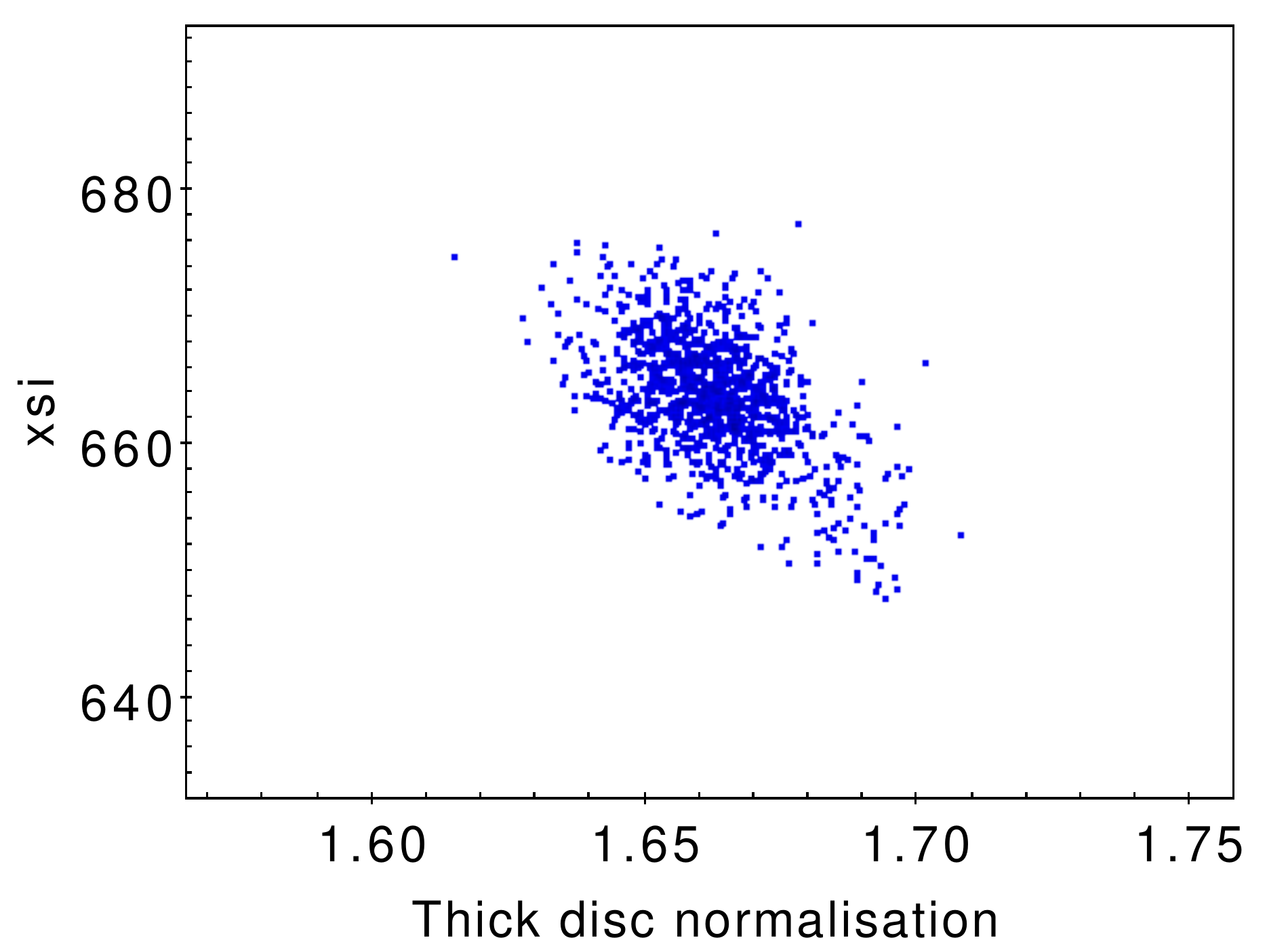}\\
    \includegraphics[width=6cm]{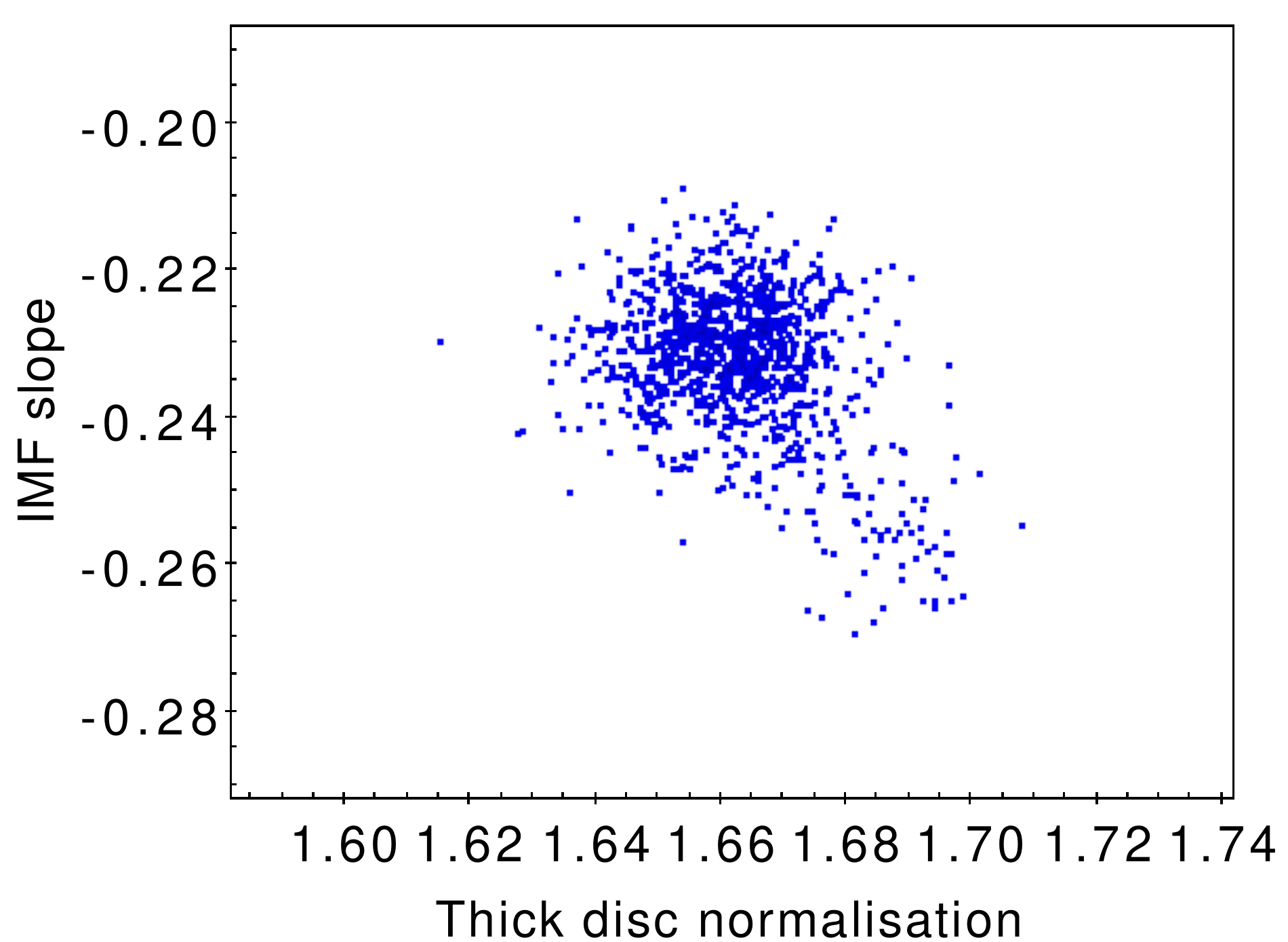}
\end{tabular}
\caption{Correlations between parameters for the best-fit model of shape A: correlations between thick-disc normalisation and other parameters (top to bottom: scale height, scale length, $\xi$, and IMF slope).}
\label{fig-correl1}

\end{figure}

\begin{figure}

\begin{tabular}{c}
   \includegraphics[width=6cm]{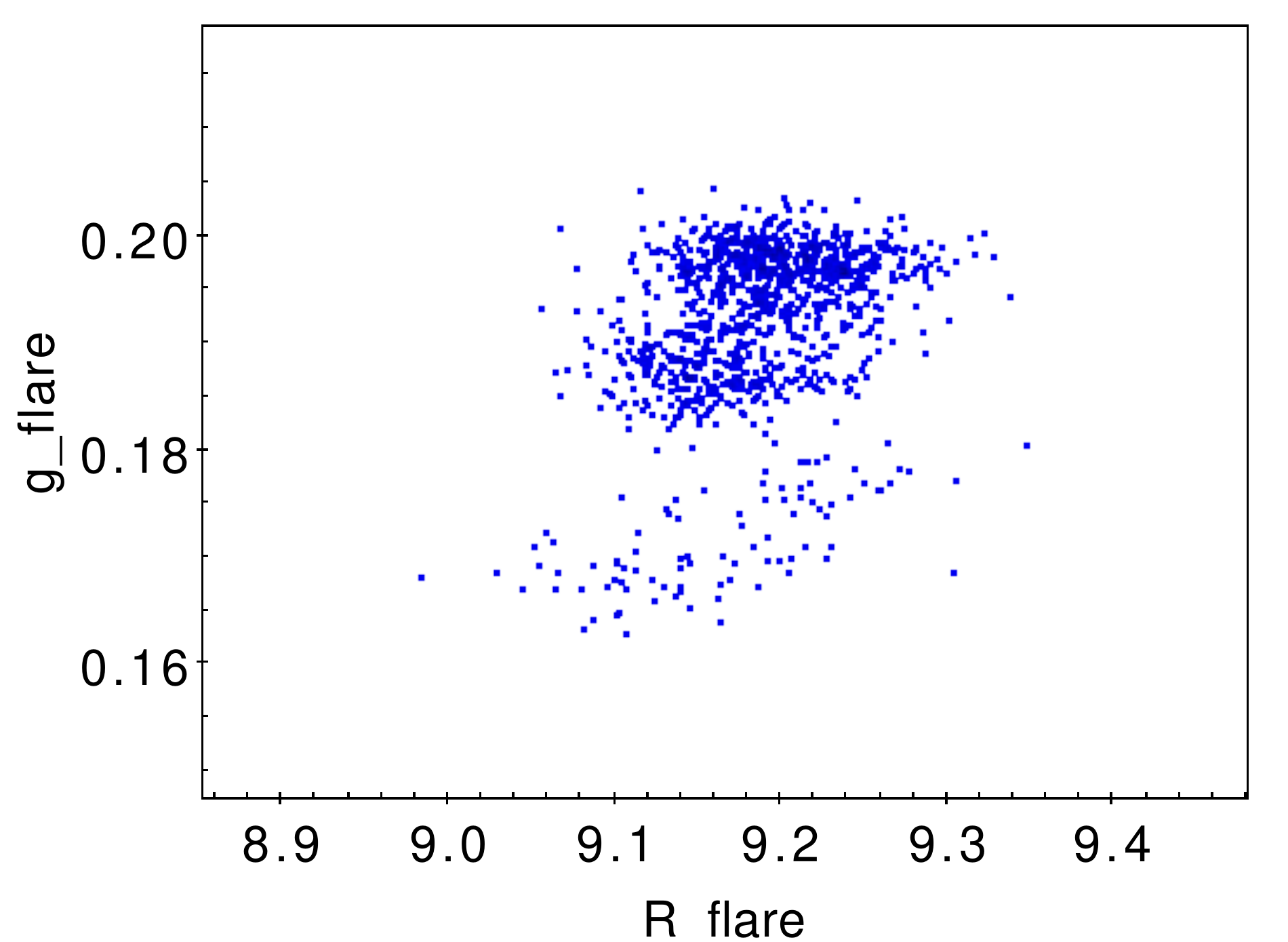} \\
   \includegraphics[width=6cm]{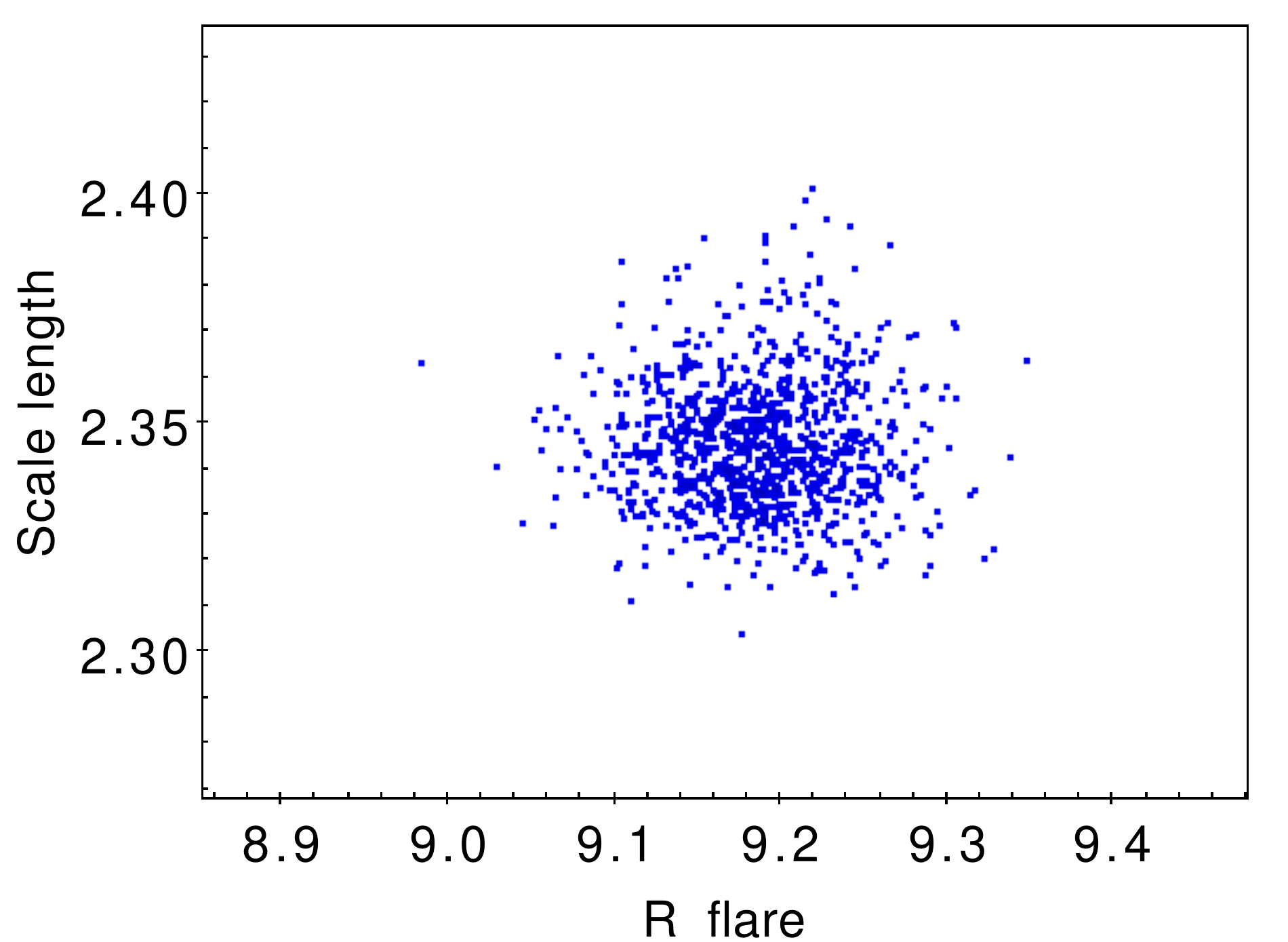} \\
    \includegraphics[width=6cm]{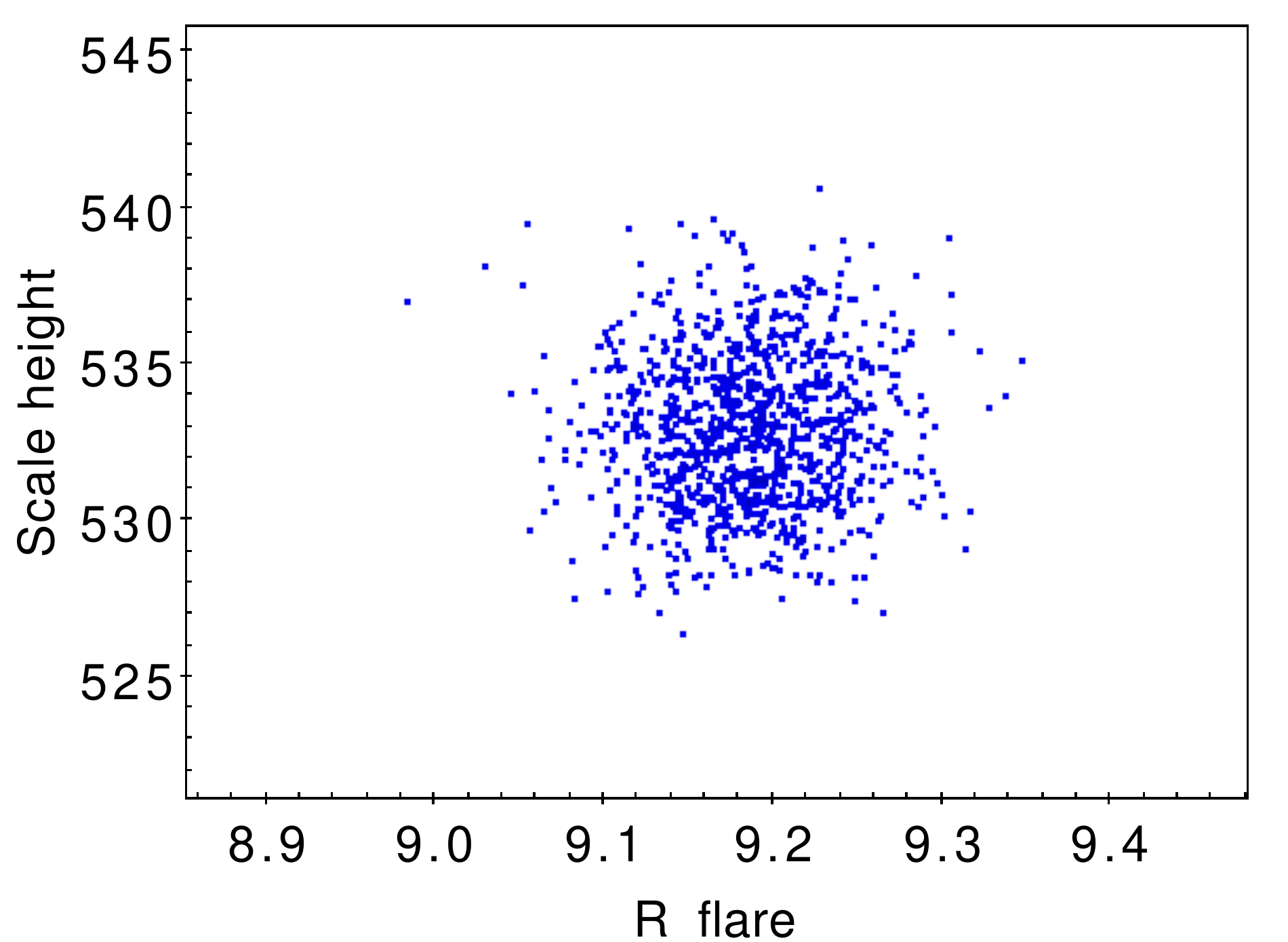}
\end{tabular}
\caption{Correlations between parameters for the best-fit model of shape A: correlations between thick-disc flare parameters: flare start radius $R_{flare}$ versus (top to bottom): flare slope $g_{flare}$, scale length, and scale height.  }
\label{fig-correl2}

\end{figure}

\begin{figure}

\begin{tabular}{c}
   \includegraphics[width=6cm]{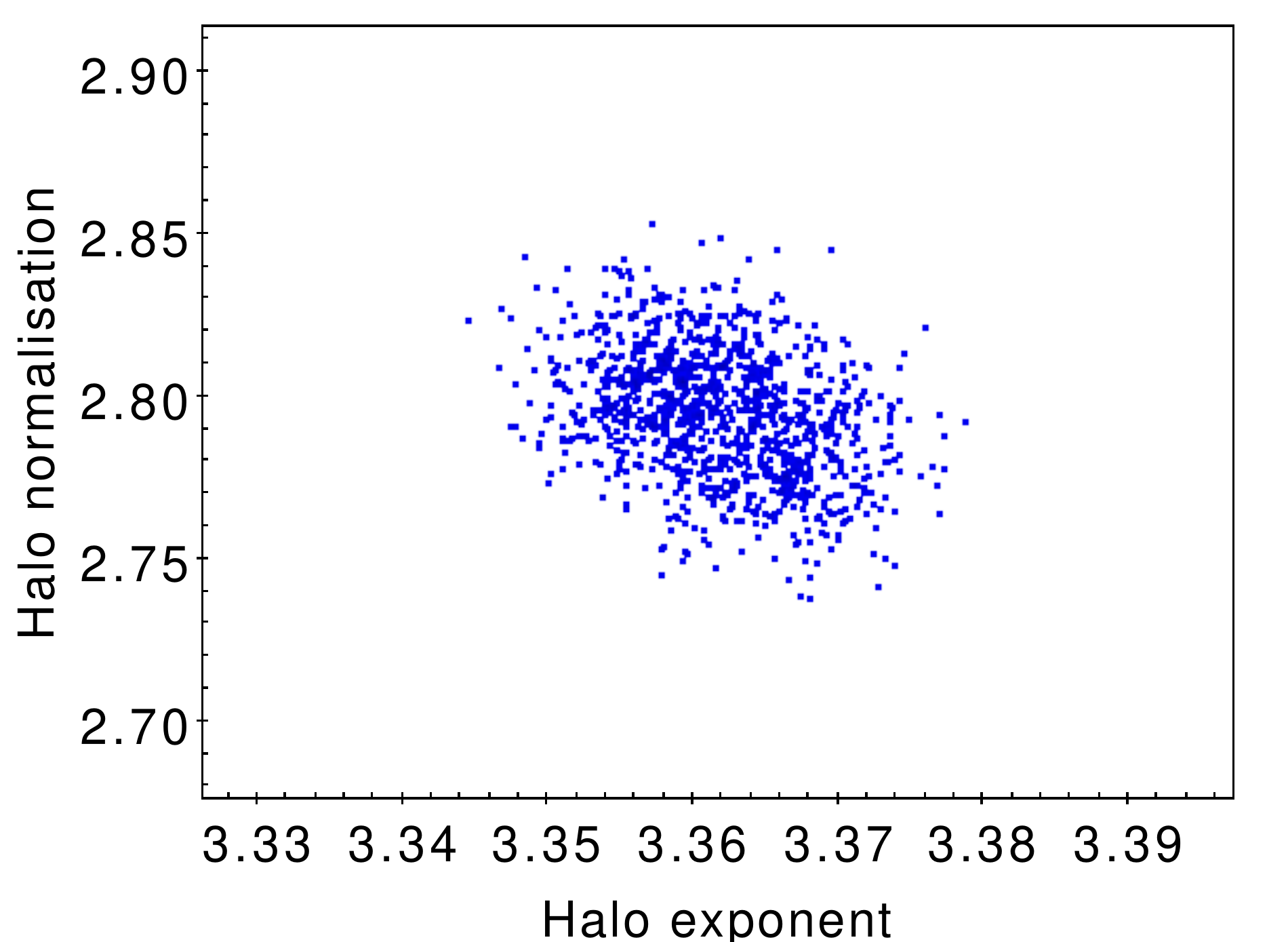} \\
   \includegraphics[width=6cm]{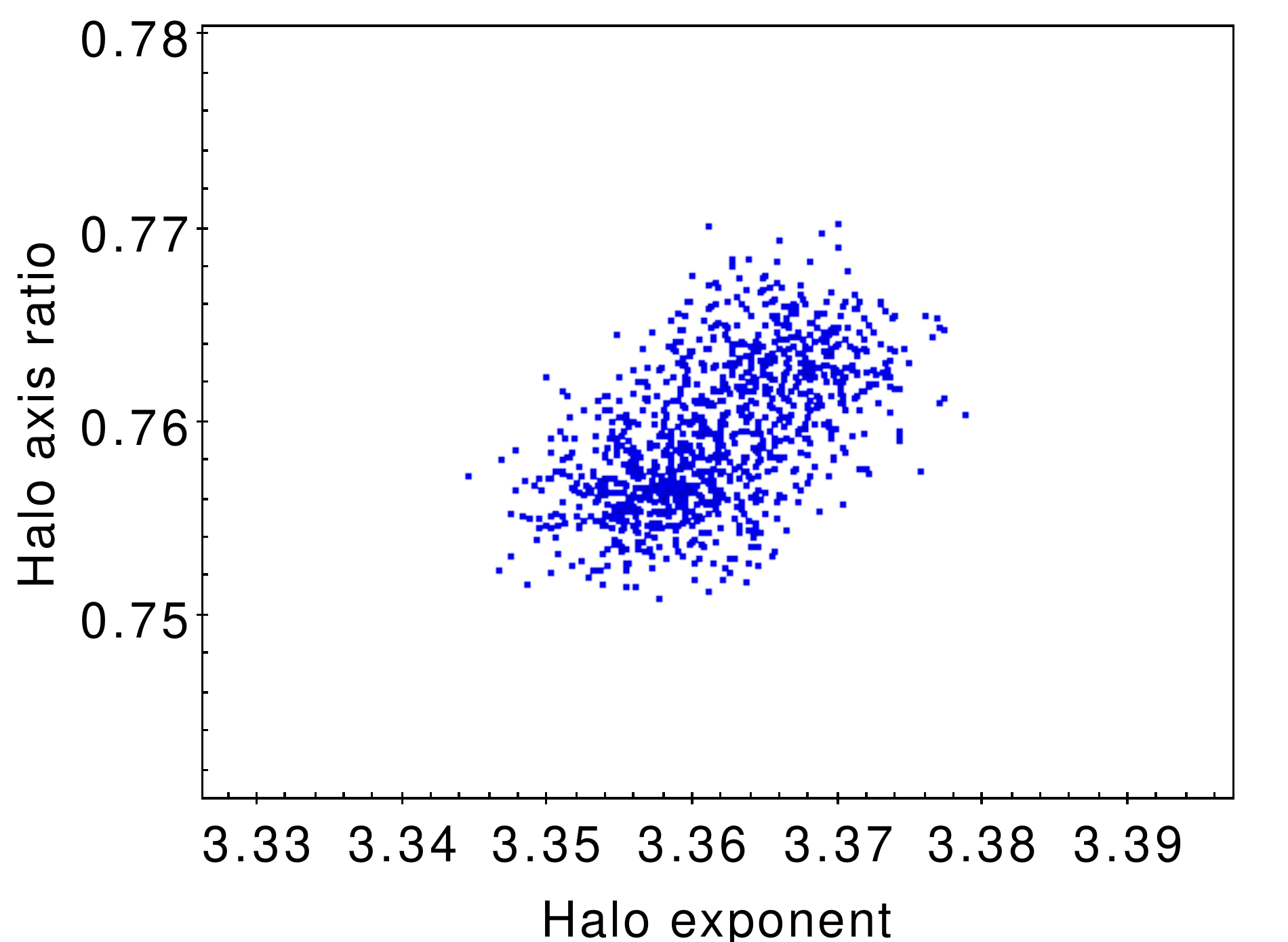} \\
    \includegraphics[width=6cm]{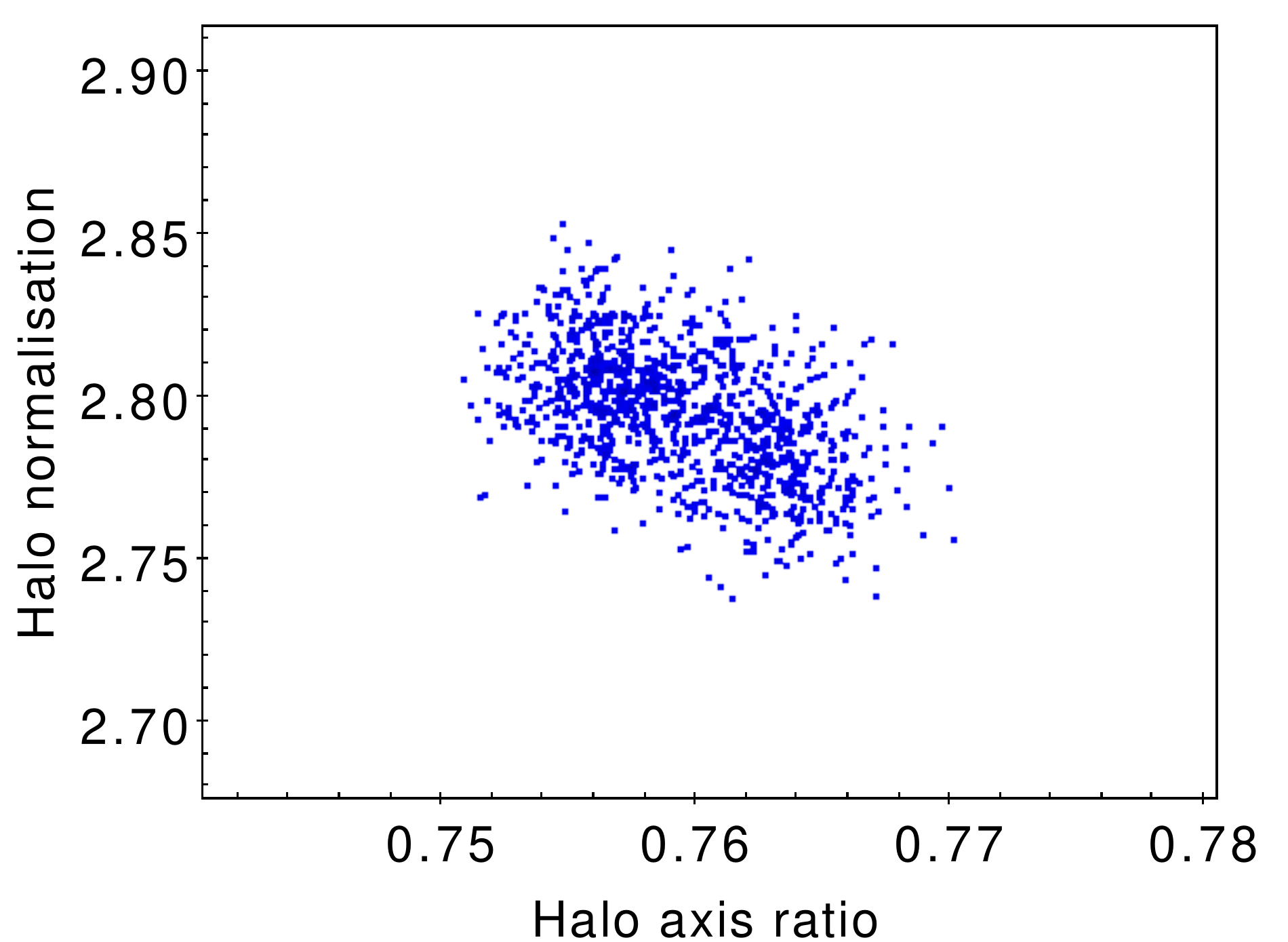}
\end{tabular}
\caption{Correlations between parameters for the best-fit model of shape A: correlations between halo parameters  (top to bottom: power-law exponent versus density normalisation, power-law exponent versus axis ratio, axis ratio versus density normalisation).}
\label{fig-correl3}

\end{figure}

We explored the correlations between parameters in fig~\ref{fig-correl1}, \ref{fig-correl2} and \ref{fig-correl3}. Each dot represents one accepted point of the Markov chain (we considered here the last third of the total chain). The correlations, if any, are weak, and the degeneracy between the thick-disc scale height and normalisation, for example, is no longer present. 
We identified a small correlation between the local density and the IMF slope of the thick disc. This is because the range of masses explored is somewhat narrow for a good constraint on the IMF slope. The local density is also very slightly correlated with the parameter $\xi$ because both parameters determine how many thick-disc stars are present at short distances from the Galactic plane. Because of the cone effect, the peak of star density occurs at some distance from the plane (depending on the scale height) and the number of star at the same time depends on the local normalisation, scale height, and $\xi$.

The parameters of the thick disc appear to be very well constrained with robust values for all independent runs we performed.
The vertical scale height is lower (535 pc) than our previous value obtained in \cite{Reyle2001} (800 pc), where we also noticed the degeneracy between the scale height and the local normalisation. In the present study these two parameters are no longer degenerate. This is because now we explore a much wider range of longitudes and latitudes which allows us to get rid of the correlation. Furthermore, in \cite{Reyle2001} we fixed the value of $\xi$ to 400 pc, while now it is a free parameter. The thick-disc normalisation has also increased compared with  \cite{Reyle2001}, but it still agrees taking into account the degeneracy in previous analyses. The scale length of the thick disc is found to be relatively short compared with previous values, 2.3 kpc, and very similar to the scale length of the thin disc (2.17 kpc) that was found in the analysis towards the Galactic centre \citep{Robin2012a}, as well as our older value of 2.2 kpc obtained from an analysis towards the Galactic anticentre \citep{Robin1992}. This means that the thick disc has a scale length of the same order as the thin disc. 

The $\xi$ parameter is found to be 	relatively high compared with the scale height. This means that the thick disc significantly deviates from a pure exponential close to the plane. The derived column density at the solar position is then higher than with an exponential law. Fig. \ref{thd-law} shows the density law we obtained in three directions. We also compare it with the best fit of the shape B.

We obtain a good constraint for the thick-disc flare,  on the Galactocentric distance of the start of the flare $R_{flare}$, but it is correlated with the flare slope. This is again an effect of the cone shape of the line of sight. This is expected because on the outskirt of the Galaxy fewer fields are used and the coverage in latitude is not complete. The maximum likelihood is found clearly, however, with a value of  $R_{flare}$ of  9.3$ \pm$ 0.3 kpc.  The flare slope does not depend on the thick-disc scale length and scale height. But when we changed the thin-disc model or the thick-disc shape (see below), the flare parameters are the least robust values and can slightly change with other assumptions. { We also rediscuss this point in sect. 7}.

The halo parameters were found completely independently of the ones of the thick disc and present no correlations with them. Within the halo parameters the correlations are weak and the constraint is very strong on the halo power-law exponent and axis ratio.

{ The accuracies of the determinated parameter values are given in the fifth column of Table ~\ref{table_t0m12b-12g3}. They were estimated from the mean of 10 runs. The dispersion on each parameter for the accepted points of the MCMC chains gives the same result.  However, these determinations can be biased because of imperfect stellar models or inadequate density laws. This is why we also explored different shapes (sect. 5.2) and different isochrones (sect. 5.3), as well as the influence of the halo shape and thin-disc models (sect. 6.) on the results.}

\subsection{Shape B thick disc of 12 Gyr and z=0.003}

\begin{figure}
\includegraphics[width=6cm,angle=-90]{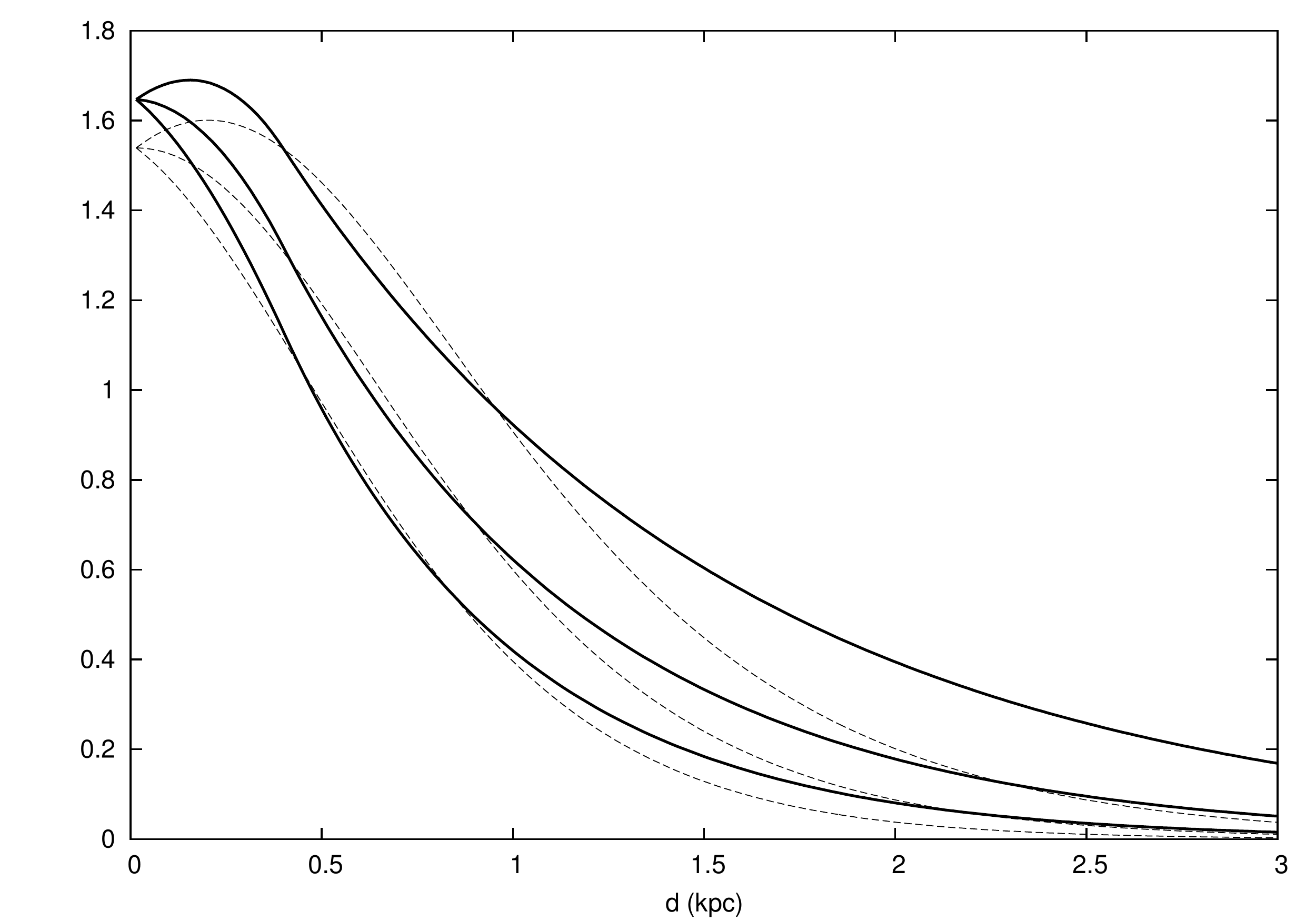}
\caption{Comparison of the best-fit thick-disc vertical density law from shape A (solid line) and shape B (dashed line) in three directions. Top lines are for l=0\degr, b=45\degr; middle lines for l=180\degr, b=45\degr; bottom lines for the North Galactic Pole. }
\label{thd-law}
\end{figure}

By turning the density shape of the thick disc to a hyperbolic secant squared (shape B) we performed a similar fit of parameters. Results are shown in Table~\ref{table-t1m12b}.
Because the number of parameters to fit is different from shape A, one has to compare the BIC value and not the likelihood.
The overall fit of the sech$^{2}$ is slightly poorer than for shape A. 
The scale height is found to be shorter in shape B than in shape A, but since the shape is different it is useful to compare the true distribution as a function of distance on the line of sight. This is shown in Fig.~\ref{thd-law} for the direction of the Galactic pole and intermediate latitudes at two longitudes. Both shape A and shape B give similar densities at distances from the plane between 0.5 and 1.5 kpc towards the Galactic pole (lowest curves). Close to the Galactic plane the sech$^{2}$ gives a slightly lower density. This is where the thin disc dominates.  At distances larger than 1.5 kpc, shape A gives slightly higher densities. Compared with previous results from \cite{Reyle2001}, the local density found here is significantly higher at short distances for both shapes, but similar at heights larger than 1.5 kpc. With shape B the flare of the thick disc is undetermined.

\begin{table}

\caption{Parameters for a thick disc of shape B: fitted model parameters, range explored (2nd and 3rd column), best-likelihood value (4th column) and standard deviation (5th column) for the thick disc of shape B, an age of 12 Gyr, and a metallicity of z=0.003. The halo is assumed to be a single power-law. BIC is -2*Lr+k*ln(n) as explained in the text.
}
\label{table-t1m12b}
\begin{tabular}{lllll}
\hline
Parameter & Min  & Max  & Best & St Dev\\
\hline
{\it Halo} \\
\hline
exponent & 2.2 & 4.5               &      3.351 & 0.01 \\
axis ratio & 0.3 & 1.2             &     0.700 & 0.01 \\
normalisation & 0. & 5.          &    3.28  & 0.01\\ 
\hline
{\it Thick disc} \\
\hline
 IMF & -1. & 1.             &   -0.32 & 0.02 \\
 scale height (pc) & 400 & 1300             &   470.0 & 0.7\\ 
 normalisation & 0. & 3.                &    1.54 & 0.03 \\
 scale length (pc) & 1500 & 5000          & 2377. & 8.2\\
 flare start radius (pc) & 7000 & 20000 &13242. & 1352.\\
 flare slope (pc/kpc) & -0.3 & 0.3 &  0. & 0.11\\
 \hline
 Lr & & & -67788. & 232. \\
 BIC & & & 135662. & \\
\hline
\end{tabular}
\end{table}

\subsection{Varying thick-disc age and metallicity }

Using our best shape (A) for the thick disc, we explored other age/metallicity combinations. Results are presented in Table~\ref{age-met}.

\begin{table*}
\caption{Best likelihood for thick-disc models with shape A with different age (in Gyr) and metallicity.}
\label{age-met}
\begin{tabular}{ccccccc}
 \hline
 z & 8  & 9 & 10 & 11 & 12 & 13\\

 \hline
0.003 & - & -69759. & -65964.& -65233. &  -64485.&-66028.    \\
0.006 &   -65922.  & -67168.&-70850. &-78596. &-84854.& - \\

 \hline
 \end{tabular}
 \end{table*}

 For a metallicity of z=0.003, the best fit is clearly obtained for an age of 12 Gyr, while for a metallicity of z=0.006 the favoured age is 8 Gyr. We did not test ages out of the range 8 to 13 Gyr, which looks reasonable from earlier studies. In all cases the z=0.003 option gives the best result. The isochrones used are imperfect stellar models. Therefore we cannot conclude about the mean age and metallicity of the thick disc with this study alone. For the metallicity, analyses of spectroscopic surveys are underway to determine it independently from the present study.  
 
 In the appendix we give examples of the quality of the fit (Fig.~A.1 to A.9)
 in different fields as star count histograms in different apparent magnitude ranges for SDSS fields and 2MASS fields. We present here the star counts for the best-fit model of shape A, as described above, as solid lines.
 {  The fit  is generally very good, apart from little differences in the shape of the blue peak, which occurs especially in the CMD at $r$ around 18.5-19 in some fields. This is where the transition occurs between the thick disc and the halo at the turnoff. The dashed lines refer to the model with a longer star formation in the thick disc, which is presented in sect.7.}

   \section{Result sensitivity to halo and thin-disc modelling}
 
 { In previous sections we explored the parameters of the thick disc, assuming a given model for the halo and the thin disc. Here we analyse whether the results obtained for the thick-disc parameters can depend on the assumptions for other populations. We first produce fits by changing the halo shapes (assuming a dual power-law as in equation~\ref{Eq-halo}, then a Hernquist halo as in eq. ~\ref{Eq-halo-Hernquist}). Furthermore we consider different models for the thin-disc population.}

 \subsection{Dual power-law halo}
 
  \begin{table}
\caption{Parameter fit for a dual halo. The 2nd and 3rd columns give the range of values explored by the ABC-MCMC method. The 4th column gives the best-likelihood value. BIC is -2*Lr+k*ln(n) as explained in the text.}
 \label{Halo-dual}
\begin{center}
\begin{tabular}{lllll}

  \hline
Parameter & Range & Range & Best value \\
& Min & Max \\
\hline
{\it Halo:}\\
Inner halo exponent & 2.2 & 4.5               &      3.393   \\
Inner halo axis ratio & 0.3 & 1.2             &     0.766     \\
Outer halo exponent &  2.2 & 4.5 &	       3.594      \\
Outer halo axis ratio &  0.3 & 1.2             &  0.766    \\
$R_{break}$ (kpc) & 25. & 150. &  49.9     \\
Inner halo normalisation & 0. & 4.         &   2.81    \\ 
\hline
{\it Thick disc:}\\
IMF slope & -1. & 1.             &  -0.25   \\
scale height (pc) & 400 & 1300             &  536.1  \\ 
 normalisation & 0. & 3.                    &  1.68    \\
scale length (pc) & 1500 & 5000          & 2369.   \\
$\xi$ (pc) & 100 & 1500                &       652.4  \\
flare start radius (pc) & 7000 & 20000 & 9256.  \\
 flare slope ($kpc^{-1}$) & -0.3 & 0.3 &   0.211 \\
 \hline
 Lr & & &  -64019.  \\
 BIC & & & 128163. \\
 \hline

 \end{tabular}
 \end{center}

 \end{table}
 
 Table~\ref{Halo-dual} shows the parameters obtained for a dual halo.  
Here the number of parameters increases to 13. 
The BIC is very similar to the reference model (very slightly better).
However, comparing different independent runs, the new parameters $a_2$, $q_2$ and $R_{break}$ are found to be poorly constrained. The break of the halo is found to occur at large distances, where the sample is rather scarce. This is probably why it did not converge well, giving best values in the range 30 and 90 kpc from run to run. This indicates that the constraint is not strong enough from the available data. The vast majority of the halo stars that enter the fit are at distances smaller than 30 kpc. The giants are present, but in smaller number and mixed with red main-sequence stars. To constrain the density of the halo at larger distance one should consider separating properly giants from dwarfs, which has not been done here where we concentrated on characterising the thick disc.
 Moreover, the exponent and flattening of the outer halo are found to be close to the inner halo. 
 The thick disc parameters are hardly affected by the halo shape because the thick disc parameters obtained in table~\ref{Halo-dual} are alike to those in table~\ref{table_t0m12b-12g3}, apart from a slight change in the flare parameters which are less well determined. 

\subsection{Hernquist halo}
 
Table~\ref{Halo-Hernquist} shows the general fit for a Hernquist halo shape. The fitted parameters are the axis ratio $q$, the exponent $d_{hern}$, the core radius $R_{core}$, and the local density. The overall fit is very good, even slightly better than the single power-law. The core radius is found to be about 2.1 kpc, but the uncertainty is large because only a few 2MASS fields in the inner Galaxy were used, and 2MASS fields are not deep enough to constrain the halo population well. The axis ratios obtained from a power law or a Hernquist law are in fact very similar, with values of 0.72 and 0.76. The exponents are different, but this is expected because of the different formulae.

 \begin{table}
\caption{Parameter fit for a Hernquist halo. The 2nd and 3rd columns give the range of values explored by the ABC-MCMC method. The 4th column indicates the best-fit parameters. BIC is -2*Lr+k*ln(n), as explained in the text.
}
\begin{center}
\begin{tabular}{lllll}
\hline
Parameter & Range & Range &  Best\\
 & Min & Max & & \\
\hline
{\it Halo:}\\
exponent $d_{hern}$ & 2.2 & 4.5               &     2.763     \\
axis ratio $q$ &  0.5 & 1.2 &	      0.774 \\
$R_{core} $(pc) & 0. & 5000.   & 2136.\\
normalisation &0. & 5.         &  2.66\\ 
\hline
{\it Thick disc:}\\
IMF slope & -1. & 1.             &    -0.22 \\
scale height (pc) & 400 & 1300             & 537.2\\ 
normalisation & 0. & 3.                  & 1.68 \\
scale length (pc) & 1500 & 5000          &   2318    \\
$\xi$ (pc) & 100 & 1500                &        647  \\
flare start radius (pc) & 7000 & 20000 &    9308.\\
flare slope (pc/kpc) & -0.3 & 0.3 &    0.22  \\
\hline
 Lr & & &      -63872.   \\
 BIC & & & 127850. \\
\hline

\end{tabular}
\end{center}
\label{Halo-Hernquist}
\end{table}%

 Note that the parameters for the thick disc are very similar to those found with the simple power-law halo, with a dual power-law halo, or a Hernquist halo, indicating that the two sets of parameters of the two populations are indeed not correlated, which ensures a robust determination of the thick disc, regardlesss of the halo shape.

\subsection{Impact of the thin-disc model on the fitted thick disc}

Recently, we have proposed an alternative modelling for the thin disc \citep{Czekaj2014}, where stars are drawn from an IMF and SFR and where binarity is properly taken into account. This model was fitted to Tycho 2 data, which cover a limited region around the Sun. It constrains the local star formation rate, which is found to decrease from 10 Gyr ago to the present time. The model has not yet been extensively compared with remote star counts, but it presents an alternative to modelling the thin disc content at the solar position. To check how the modelling of the thin disc can impact the fit of the thick disc, we considered the two models A and B proposed by \cite{Czekaj2014} and combined them with the thick disc and halo populations. 
Model A and model B differ by their local density (0.044 \Msun pc$^{-3}$ for the former, 0.039 for the latter), and by the assumed IMF (Haywood et al, (1997) for model A and a combination of Kroupa (2008) at mass larger than 1.53 \Msun and Haywood et al. (1997) at lower mass).

Then we fitted the thick disc and halo as before. In Table~\ref{mevol} we present the parameters of the thick disc obtained using either of these two new models and compare them with the standard BGM used in the previous section. It is clear that the fit of the thick disc is very little affected by the thin disc. The only parameter that changes significantly is the thick-disc flare. But this is in the case of model A, which also gives the poorest likelihood. We conclude that the major thick-disc parameters are robust against the hypothesis adopted about the thin-disc population.

\begin{table}
\caption{Comparison of thick-disc parameters obtained with various thin disc models.  We considered model A and B from Czekaj et al. (2014), which differ by their IMF and local density. The 4th column recalls the values obtained earlier by the standard BGM.%
}
\begin{center}
\begin{tabular}{lllll}
\hline
Parameter & model A & model B & Reference model  \\
\hline
{\it Halo:}\\
power law exponent & 3.11 &  3.08 &  3.36 \\
axis ratio $q$ & 0.582 & 0.844 & 0.757  \\
normalisation & 1.000 & 1.69 &  2.82\\
\hline
{\it Thick disc:}\\
IMF slope & -0.95 & -0.86 & -0.22\\
scale height (pc) & 549. & 596.1 & 535.2\\ 
normalisation & 1.52 & 1.53 & 1.647\\
scale length (pc) & 2410.  & 2431. & 2362. \\
$\xi$ (pc) & 779. & 675. & 662.8\\
flare start radius (pc) & 15034. & 9188. & 9261.\\
flare slope (pc/kpc) & -0.009 & 0.189 & 0.1310 \\
\hline
 Lr & -81241. &     -73707.  & -64485. \\
\hline

\end{tabular}
\end{center}
\label{mevol}
\end{table}%

 \section{ Duration of star formation in the thick-disc phase}
 
{  In the results above we considered a single age and single metallicity for the whole thick disc. Most recently, several studies \citep{Bovy2012a,Haywood2013} claimed that the thick disc formed over a longer period and that there is a continuity of star formation between the thick disc and the thin disc.  We attempt to simulate this by considering two episodes of formation at a time interval of 1 or 2 Gyr. The impact on changing the age by 1 Gyr is small but not negligible on the isochrone, and the position of the turnoff is changed such that the CMDs can be better or poorer fitted with different ages. But the effect is not strong enough to cause a difference between the sum of two episodes at 1-2 Gyr intervals and a continuous star formation during such a time lap. This is mainly because the observed turnoff position in CMDs is widened by photometric errors on the one hand and by the metallicity dispersion on the other.
Therefore we consider that using two isochrones of different ages is a simplified approach to simulate a continuity between the thin and thick disc. We performed this test with different ages and a fixed metallicity of z=0.003 for the two episodes of thick-disc formation. To avoid too many parameters to fit we here assumed a shape B thick disc (sech$^2$), which gives similar results to shape A, but with two parameter less. The result of the test is given in Table~\ref{table-2thd}.  
 
 \begin{table}

\caption{Test with two thick-disc episodes. Parameters are compared for a single episode (2nd column), two episodes with different ages (columns 3 to 5) obtained with a free Hernquist halo and a thick disc with shape B (sech$^2$). For each parameter the first line is for the older thick-disc episode and the second line for the younger thick-disc episode.}
\label{table-2thd}
\begin{tabular}{lllll}
\hline
Old thick disc &  12 Gyr  & 12 Gyr   &  12 Gyr & 11 Gyr  \\
Young thick disc & - & 11 Gyr & 10 Gyr & 10 Gyr    \\
\hline
scale height (pc) &465.  & 826. & 795.  & 824.\\ 
	& - & 359. & 345. & 348.\\\\
	
scale length (pc) & 2305.  & 3077. & 2919. & 2907 \\
	& - & 1986. & 2040. & 2089. \\\\
normalisation & 1.55  & 0.21 & 0.25 &0.19 \\
	& - & 1.54 & 1.63 & 1.65  \\\\
 flare start radius (pc) & 9359. &  10020. & 9543 & 9757 \\
 	& - & 17364 & 15340. & 14400.\\\\
flare slope (pc/kpc) & 0.187  & 0.09  & 0.06 & 0.09\\
	& -  & 0.02 & 0.06 & -0.08\\\\
 Lr & -66085. &  -60360. & -59077  & -61015 \\
 BIC & 132035 & 120566.  &  118000. & 121876\\
\hline
\end{tabular}
\end{table}

This table shows systematic features that are particularly interesting for understanding the thick-disc formation and its relation with the thin disc. In the three cases (we also performed other tests that yielded the same systematics; these are  not shown here) the likelihood (and the BIC) is better than for a single-age thick-disc episode. The systematics are that the older thick disc has a larger scale height and scale length, and is less populated than the younger thick disc. In addition the flare is only present in the old thick disc.

To understand the changes in the simulated CMDs when the thick disc is
simulated by a short period of formation or a longer period
(here simulated by two successive star formation episodes), we have plotted the star count residuals in the CMDs in different fields (number of stars observed in each colour/magnitude bins minus simulated stars). This is presented  in Fig.~\ref{cmd-2m-1thd} and \ref{cmd-sdss-1thd} for a single episode and 
in Fig.~\ref{cmd-2m-2thd} and \ref{cmd-sdss-2thd} for two episodes, each for nine 2MASS fields and for nine  SDSS fields at various longitudes and latitudes. In the appendix we also compare the histograms of simulated star counts in magnitude bins with data, considering the model with a single star formation episode and the model with two episodes. As for 2MASS CMDs and star counts, there is no significant difference in the two cases. But in SDSS fields, for a single episode the CMD residuals in the data are higher than in the model, so that the model does not reproduce {the turnoff part of the CMD well}. When the second episode is introduced, the CMD is reproduced better. The main difference arises at magnitudes 17-19 at g-r around 0.4{, at the turnoff. In the histograms in the appendix the difference is clear in Fig. A.1 and A.3 for example on the turnoff at magnitudes 17-18, or in Fig. A.4 at magnitude 18-19. There is no other significant difference between the two models but the turnoff}. This is a good indication that the thick-disc formation probably has occurred during a longer period.

\begin{figure*}
\begin{center}
\includegraphics[width=14cm]{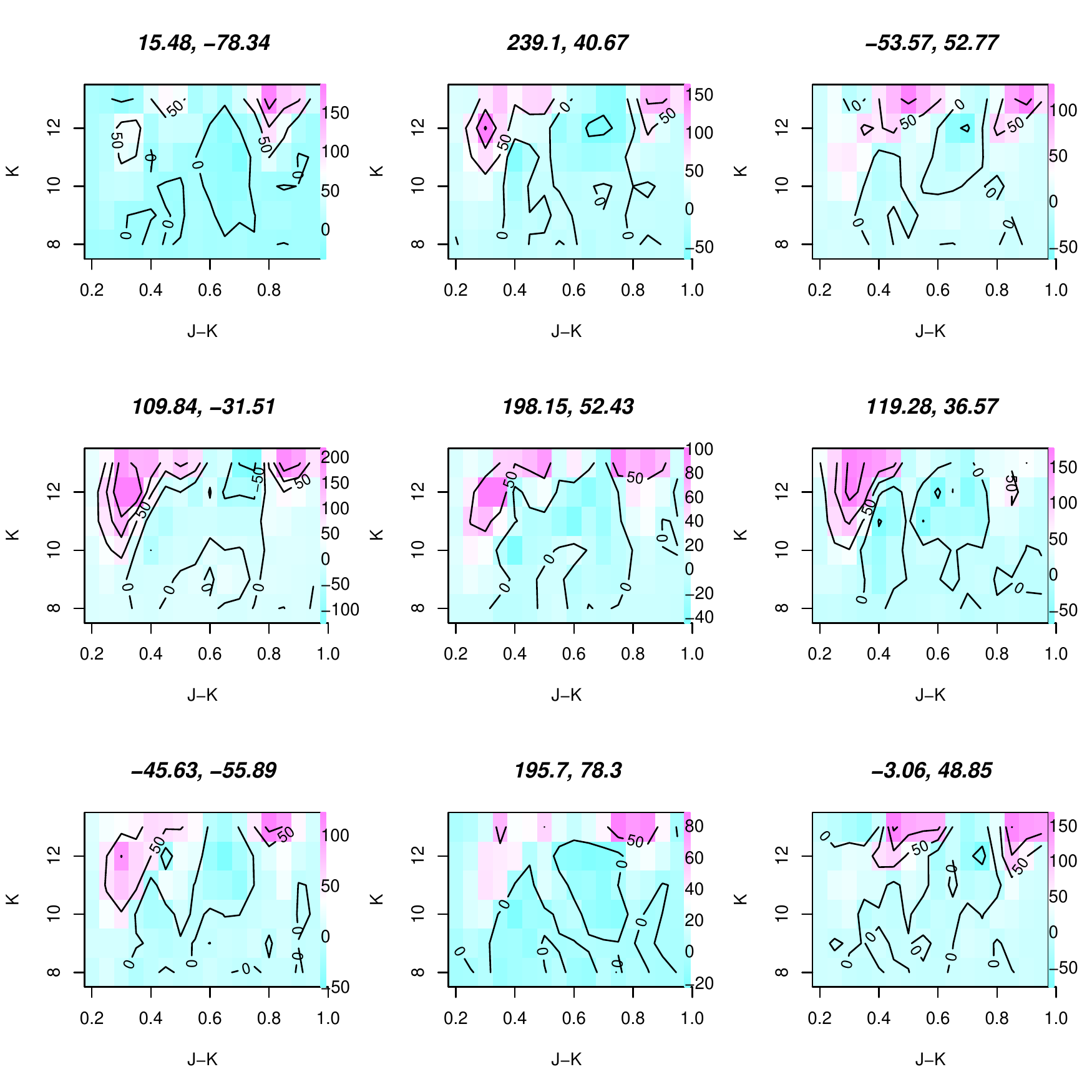} 
\caption{Residuals of star density in colour magnitude diagrams for 9 2MASS fields for the model with a single star formation episode. The longitude and latitude of each field considered are indicated in each panel. The residuals are indicated in the colour level and in contours that are labelled with their level.}
  \label{cmd-2m-1thd}
\end{center}
\end{figure*}

\begin{figure*}
\begin{center}
\includegraphics[width=14cm]{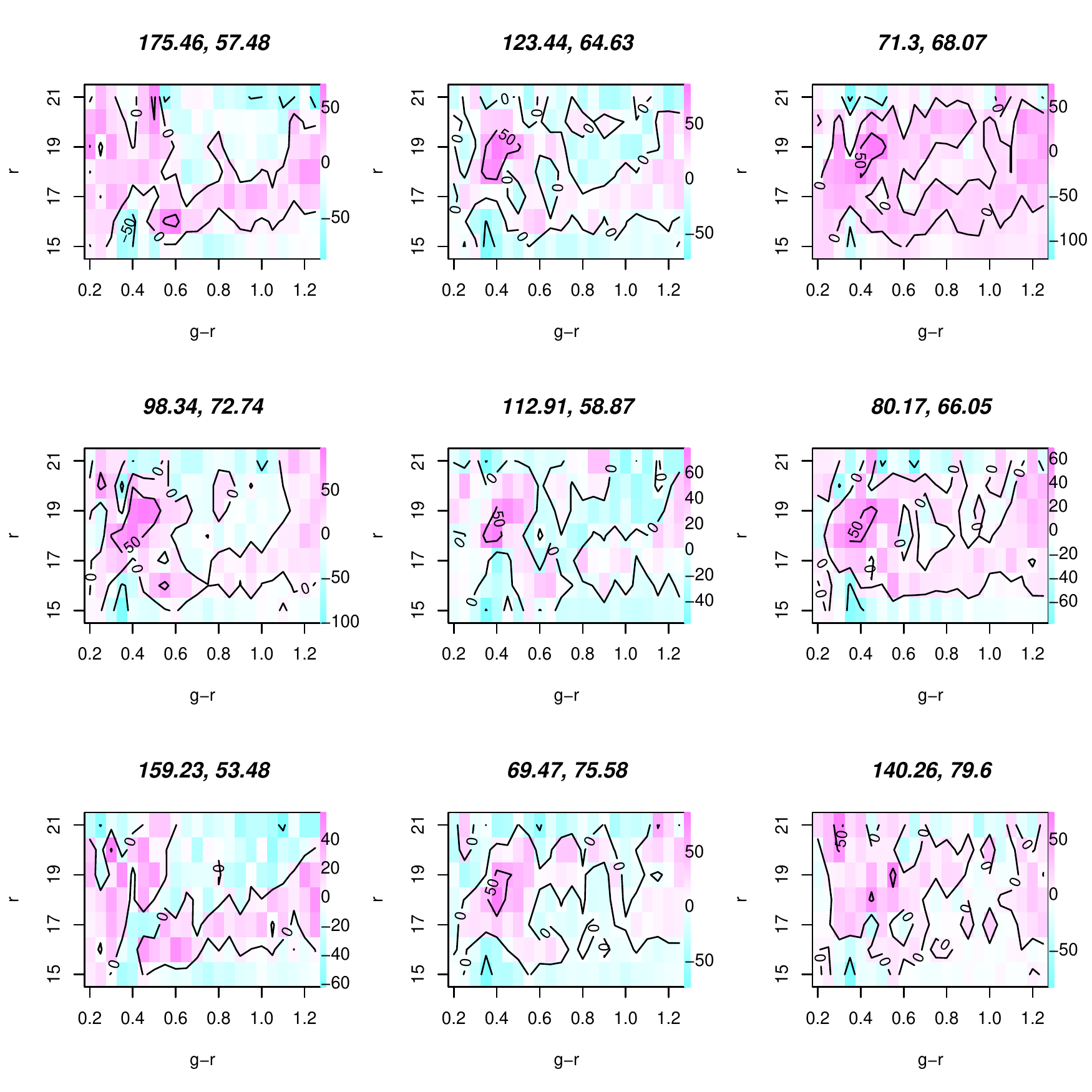} 
\caption{Residuals of star density in colour magnitude diagrams for 9 SDSS fields for the model with a single star formation episode. The longitude and latitude of each field considered are indicated in each panel. The residuals are indicated in the colour level and in contours that are labelled with their level.}
\label{cmd-sdss-1thd}
\end{center}
\end{figure*}

\begin{figure*}   
\begin{center}
\includegraphics[width=14cm]{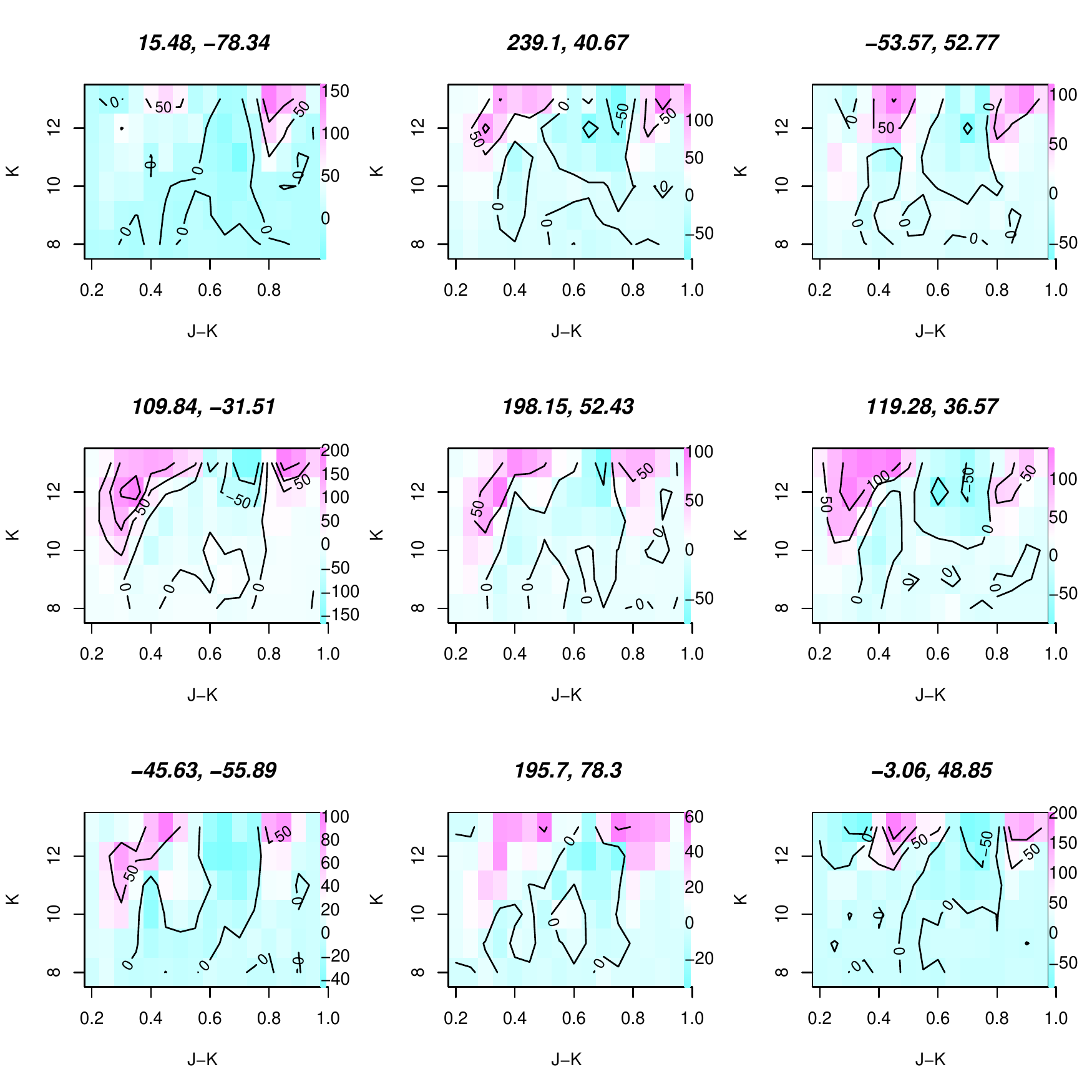} 
\caption{Residuals of star density in colour magnitude diagrams for 9 2MASS fields for the model with two episodes of star formation in the thick disc. The longitude and latitude of each field considered are indicated in each panel. The residuals are indicated in the colour level and in contours that are labelled with their level.}
\label{cmd-2m-2thd}
\end{center}
\end{figure*}

\begin{figure*}
\begin{center}
\includegraphics[width=14cm]{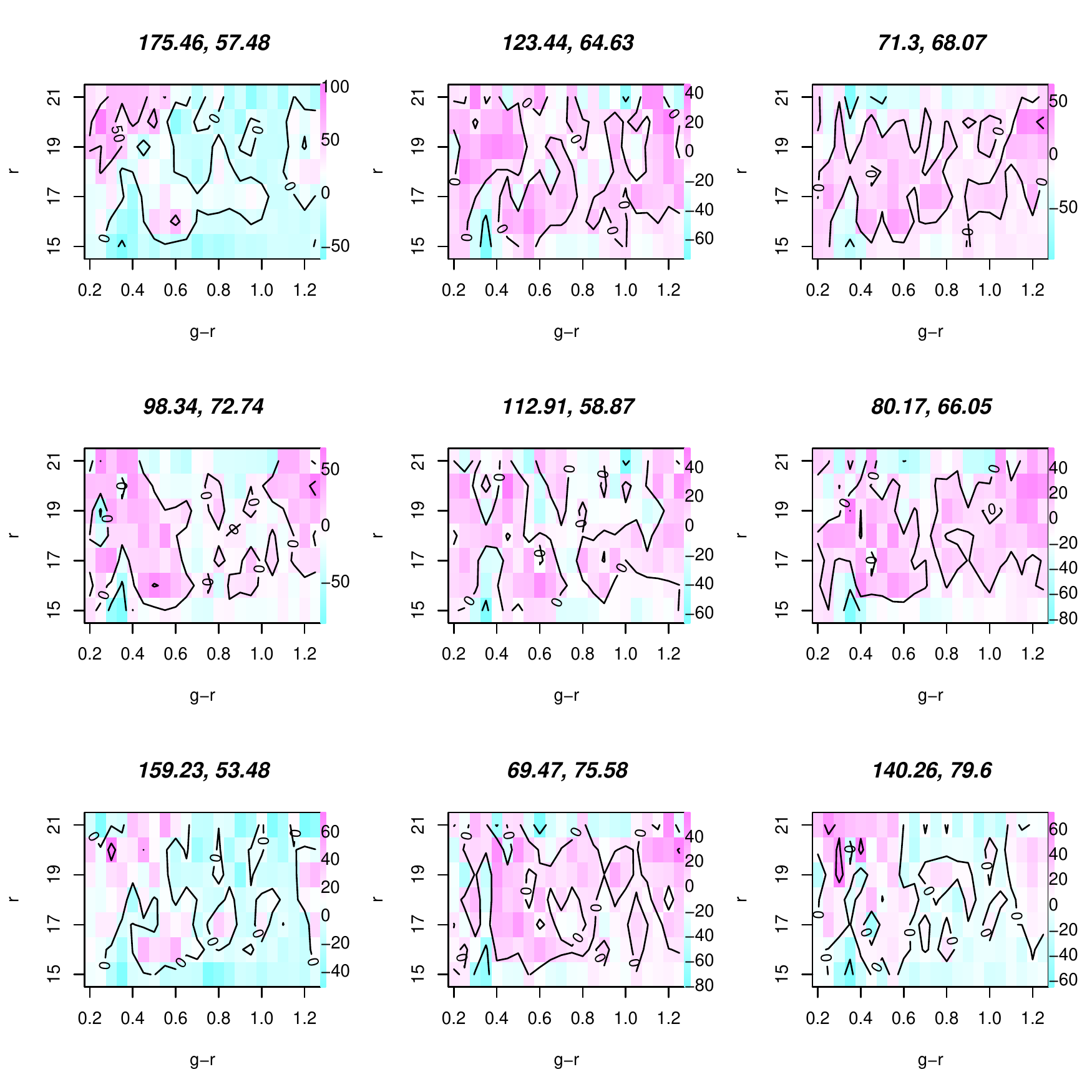} 
\caption{Residuals of star density in colour magnitude diagrams for 9 SDSS fields for the model with two episodes of star formation in the thick disc. The longitude and latitude of each field considered are indicated in each panel. The residuals are indicated in the colour level and in contours that are labelled with their level.}
\label{cmd-sdss-2thd}
\end{center}
\end{figure*}

In Fig.~\ref{hz-hr} we compare the parameters of the thick disc as a whole with the parameters for the young thick disc and the old thick disc. The normalisations show that the younger thick disc dominates in density. This explains why the parameters are similar to those of the younger thick disc when it is considered as a single age population. The important result is that the scale height and scale length are both larger for the old population than for the young population. This clearly indicates a contraction of the galaxy during the thick-disc formation. When we also consider the flare, it is strongly marked in the old thick disc but not in the young thick disc. 

\begin{figure}
\begin{center}
\includegraphics[width=12cm, angle=-90]{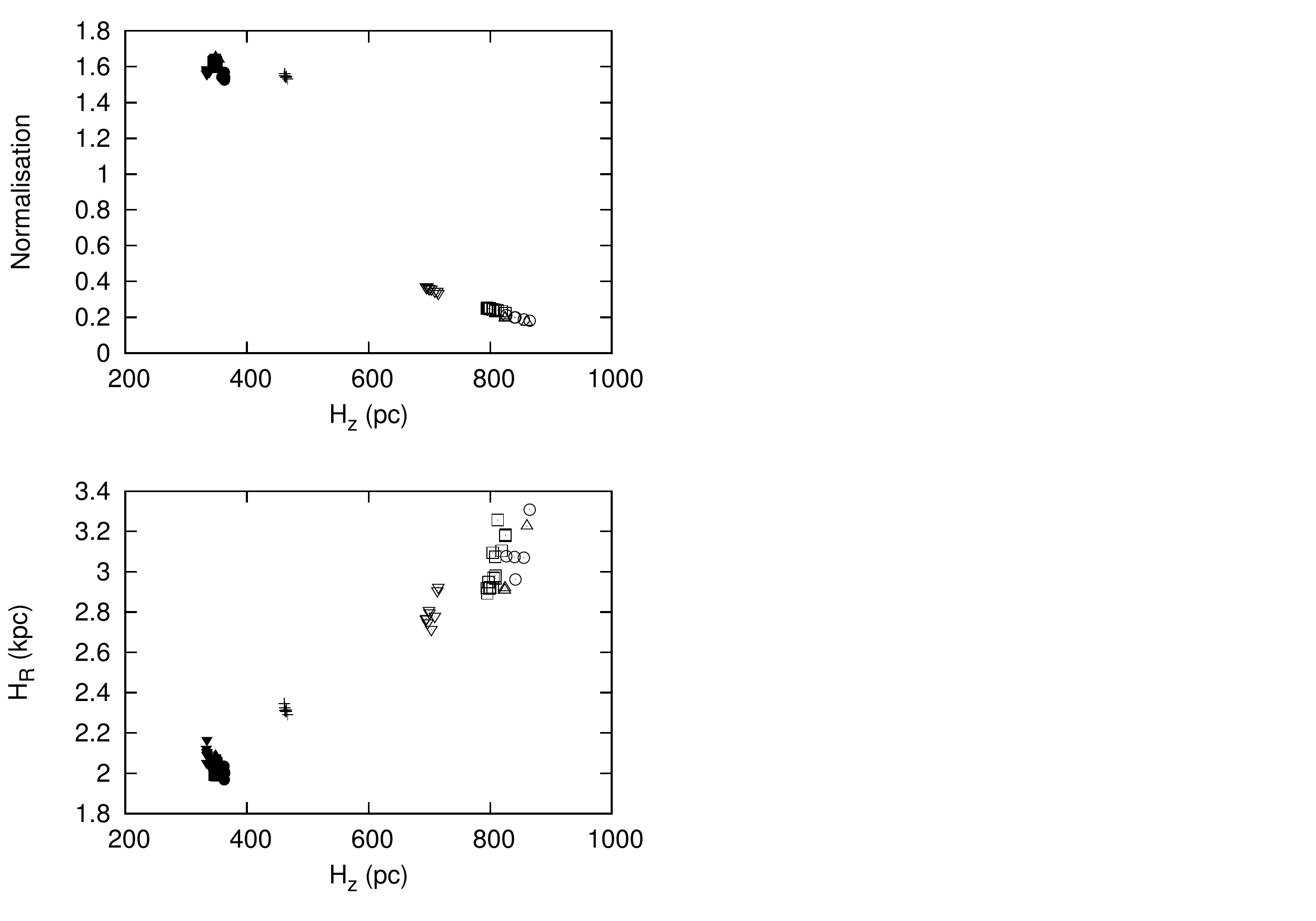}
\caption{Comparison of the thick-disc parameters as a whole with parameters for the young and the old thick discs. Open symbols are for the old thick disc (with either 12 or 11 Gyr). Full symbols are for the young thick disc (with either 11 or 10 Gyr). Plus signs are for the thick disc with a single age. }
\label{hz-hr}
\end{center}
\end{figure}

We summarise our results below, before discussing the implications. 

\begin{enumerate}
\item The thick disc is best simulated with an {extended period of star formation}. During its formation, it has undergone a contraction since the older part of the thick disc has a longer scale length and a higher scale height than the younger part.  The second  (younger) episode of thick-disc formation is found with a scale height of about 340 pc, which is twice that of the oldest part of our thin disc\footnote{In our model the old thin-disc scale height is constrained by other studies to have an eccentricity of 0.0677, which in practice is close to a $sech^2$ with a scale height of 170 pc, or to an exponential with a scale height of about 200 pc.}. 
\item The stellar mass is smaller in the old episode than in the young episode. This means that most of the stars were formed during the second episode or at least 11 Gyr ago.
\item {A significant flare appears in the early thick disc, but is not found in the younger thick disc. In the main episode of thick-disc formation, we see no evidence of a flare, at least up to a Galactocentric distance of 13 kpc. This might be because
 our study did not include enough fields in the anticentre direction at low latitudes, but more probably, it is a specific feature of the thick disc.}

\end{enumerate}

We discuss these results in the light of scenarios and simulations for thick-disc formation in Sect.~\ref{Discussion}.
}

 \section{Inner thick disc and the bulge counterpart}
 
 In \cite{Robin2012a} we investigated the inner Galaxy and discovered that the projected stellar density in the bulge region is better reproduced by the population synthesis model if we assume that the bar and bulge are two distinct populations. We fitted these two populations and found a bulge with a scale height of the order of ~800 pc and that extended radially at least to 2.5 kpc. We discussed that this bulge population could be either a small classical bulge or be related to the local thick disc, although we had no constraints on its age.
 In this study we assumed a thick disc with a scale length of 2.5 kpc and a scale height of 800 pc, following \cite{Reyle2001}.
 
The new analysis presented here provides a very good constraint on the thick-disc scale length which is found to be 2.3 kpc for the single-burst hypothesis, or ranging from 3 kpc to 2 kpc in the longer star formation hypothesis. The change of scale length seems small, but when one changes the scale length from 2.5 kpc to 2.3 kpc, the density at the Galactic centre increases by 32\%. We now have to reconsider the effect on this new thick disc in the inner Galaxy.
  
 \subsection{Star density in the bulge region}
 
\begin{figure}[htbp]
\begin{center}
\includegraphics[width=6cm,angle=-90]{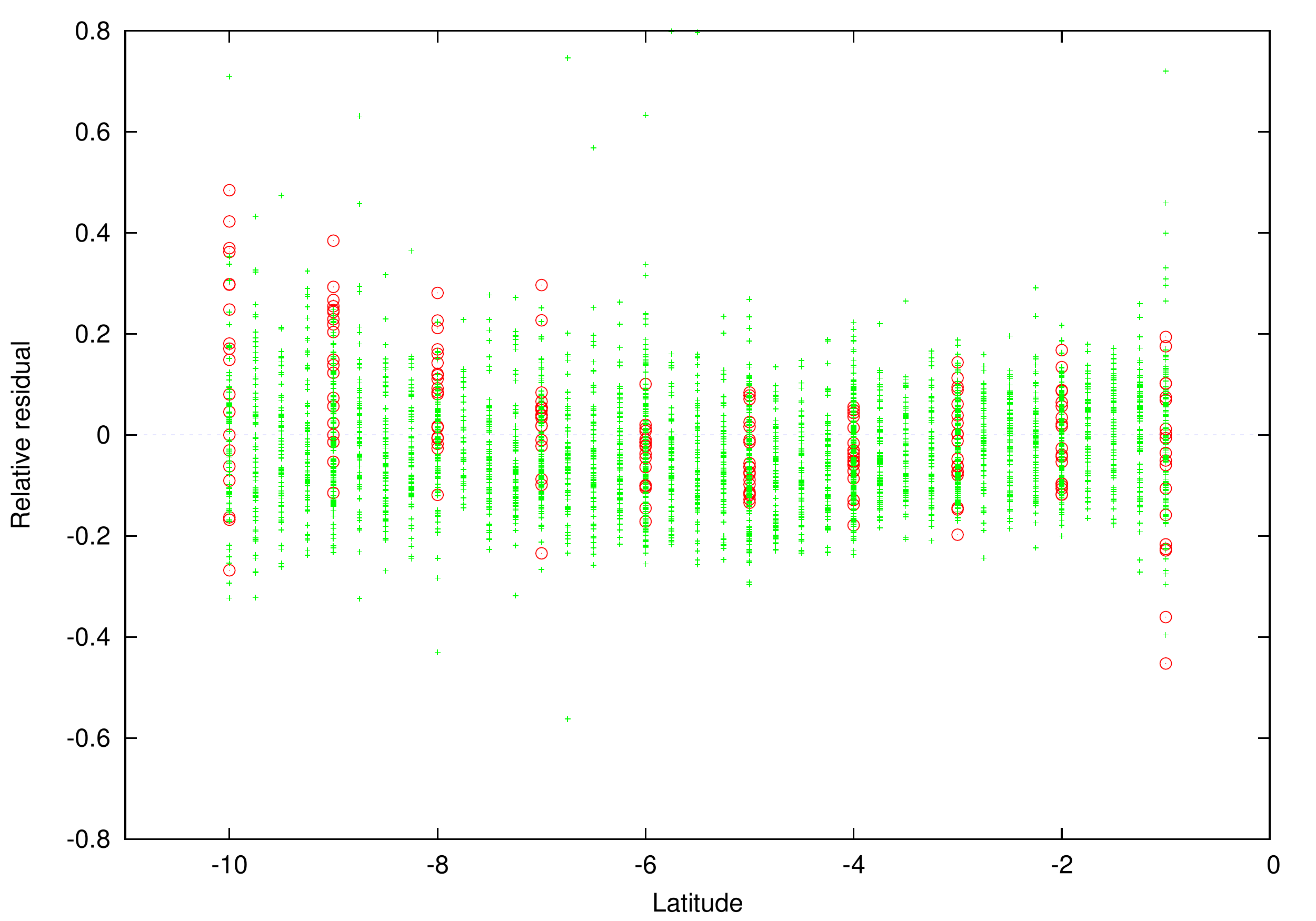}
\includegraphics[width=6cm,angle=-90]{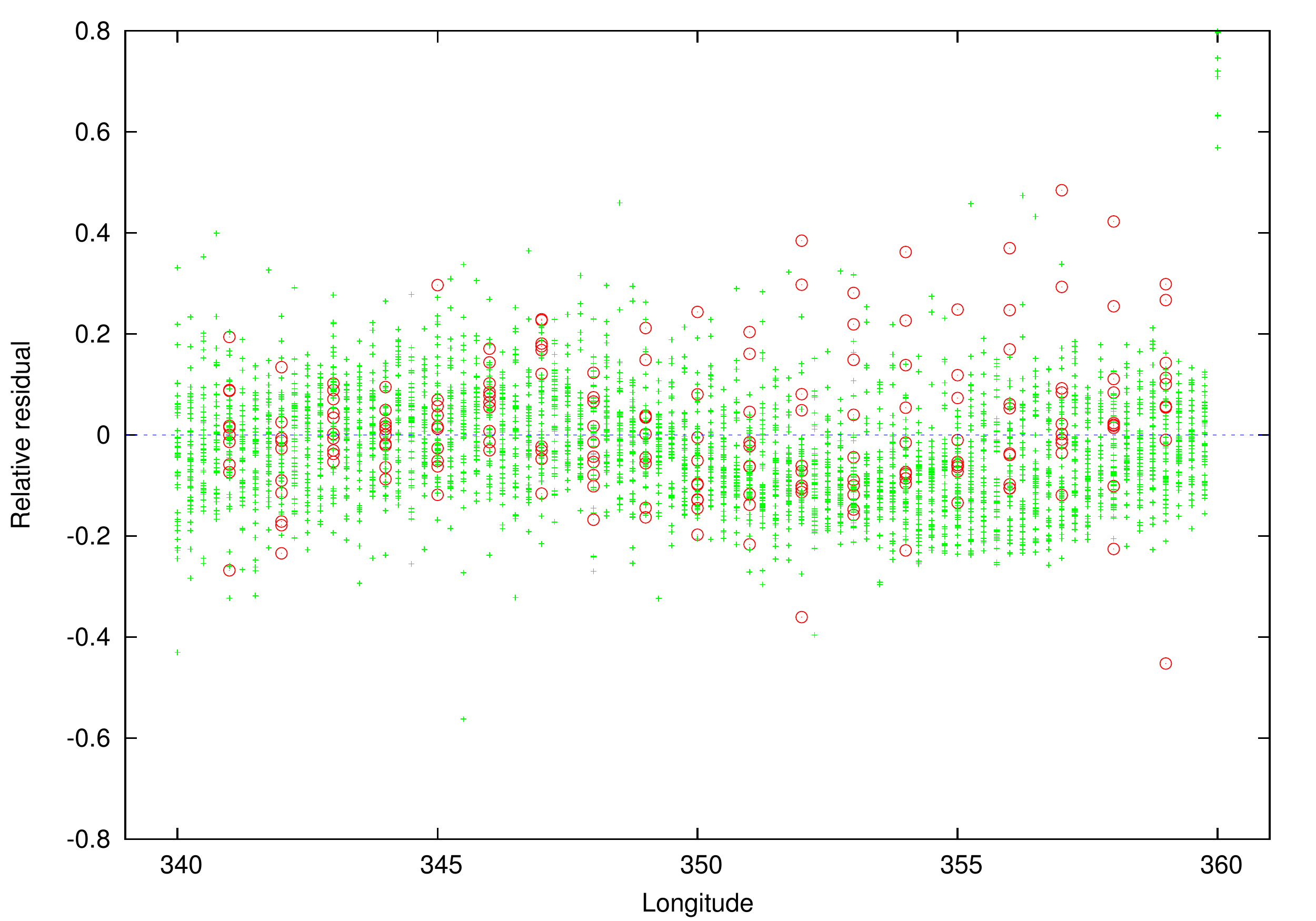}
\caption{Relative simulation residuals with regard to 2MASS star counts in the bulge region, assuming the single-burst thick disc. Top panel: relative residuals as a function of latitude. Bottom panel: the same as a function of longitude. Red open circles: residuals with the bulge model from Robin et al. (2012) assuming a small classical bulge. Green crosses: model with the revised thick disc and no classical bulge. }
\label{bulge-1thd}
\end{center}
\end{figure}

\begin{figure}[htbp]
\begin{center}
\includegraphics[width=6cm,angle=-90]{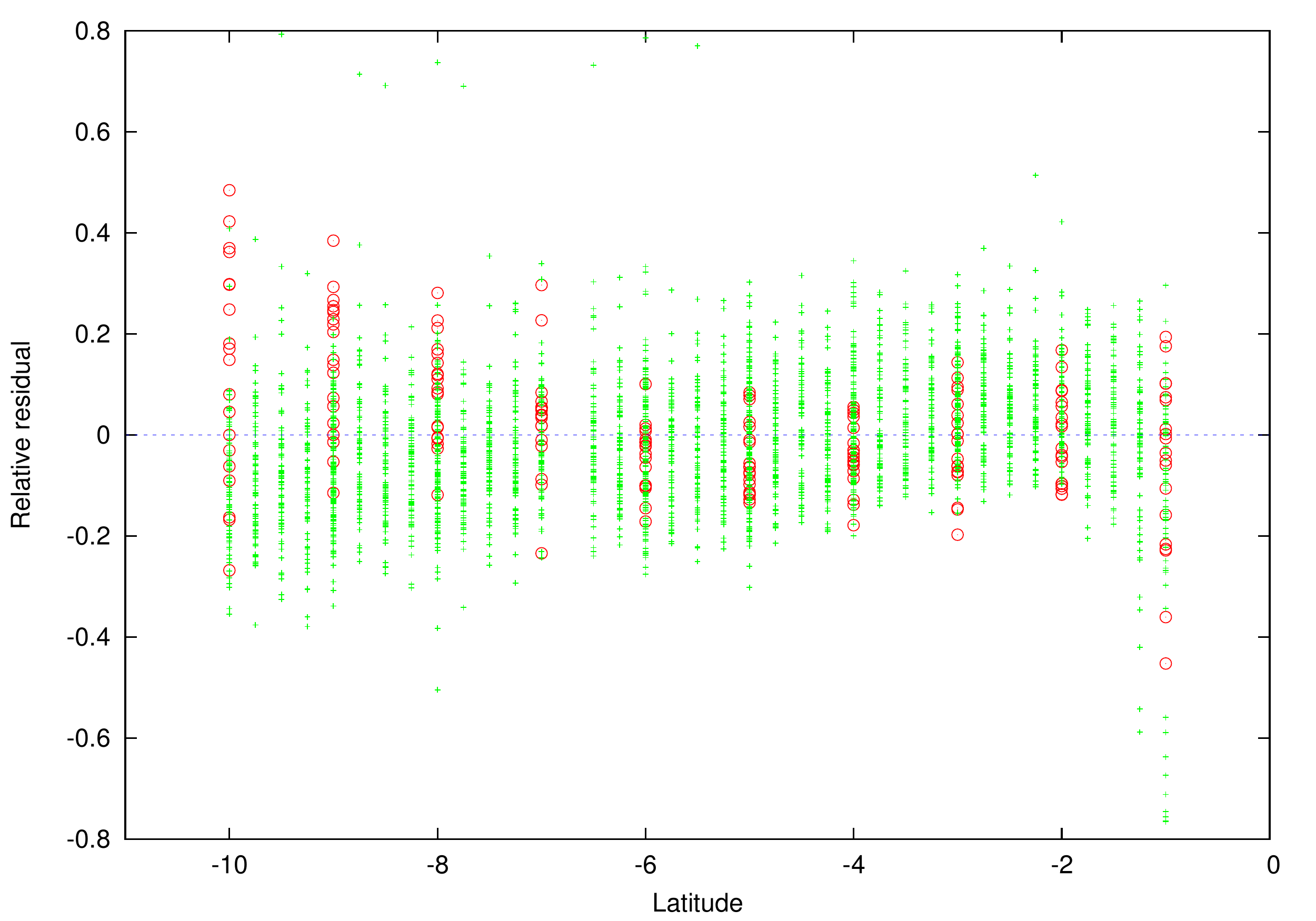}
\includegraphics[width=6cm,angle=-90]{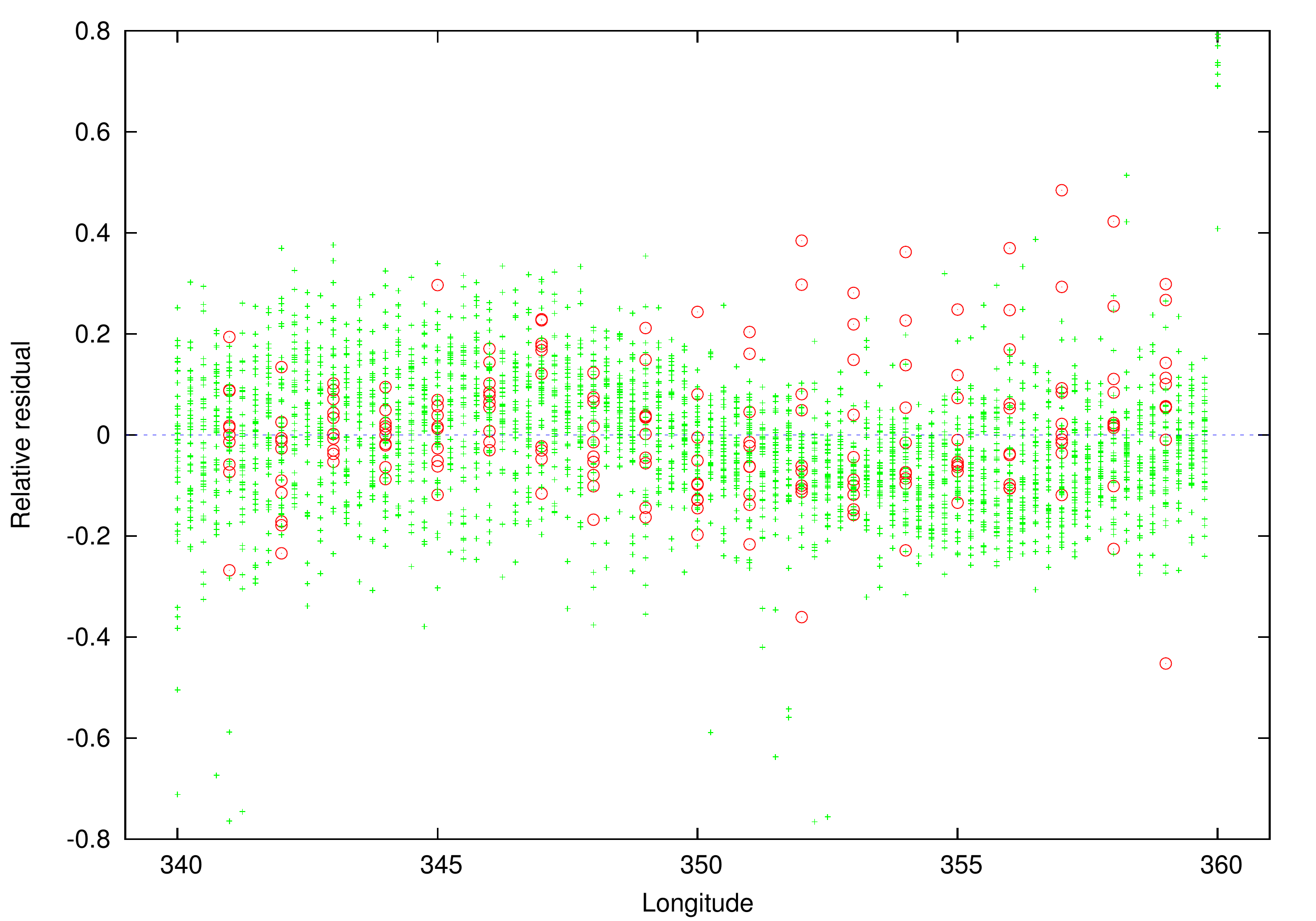}
\caption{Relative simulation residuals with regard to 2MASS star counts in the bulge region,  assuming the longer star forming thick disc. Top panel: relative residuals as a function of latitude. Bottom panel: the same as a function of longitude. Red open circles: residuals with the bulge model from Robin et al. (2012) assuming a small classical bulge. Green crosses: model with the revised thick disc and no classical bulge. }
\label{bulge-2thd}
\end{center}
\end{figure}

 We recomputed the bulge region star counts using the method reported in \cite{Robin2012a} and compared the relative residuals ($N_{Model}-N_{Obs})/N_{Obs}$  in the bulge region for a single burst or a longer star formation for the thick disc. 
 The included populations are the thin disc, peanut-shaped bar, thick disc, and stellar halo (using the parameters from this study). { 
 The results are presented in Fig.~\ref{bulge-1thd} and \ref{bulge-2thd}. We see that with the new thick disc (and no classical bulge population) the model counts reproduce the 2MASS star counts in the latitude range [-10:10] and longitude range [-20:20] very well: the new thick disc (either version) extrapolated into the bulge region gives similar (or even better) residuals as does summing the old thick disc and thick bulge which was found in the 2012 model. The version with a longer star formation period gives  slightly lower residuals than the single burst. We note that some systematics seem to appear especially at longitudes close to the Galactic centre. We plan to revisit the fit of the bar using this new thick-disc  model to take  these systematics into account.}
 
With this new thick disc, there is no more need for a classical bulge population in the inner galaxy, while the bar (or pseudo-bulge) population with its peanut shape remains nearly unchanged.
 
 \subsection{Implication for the inner rotation curve}

\begin{figure}
\includegraphics[width=8cm,angle=0]{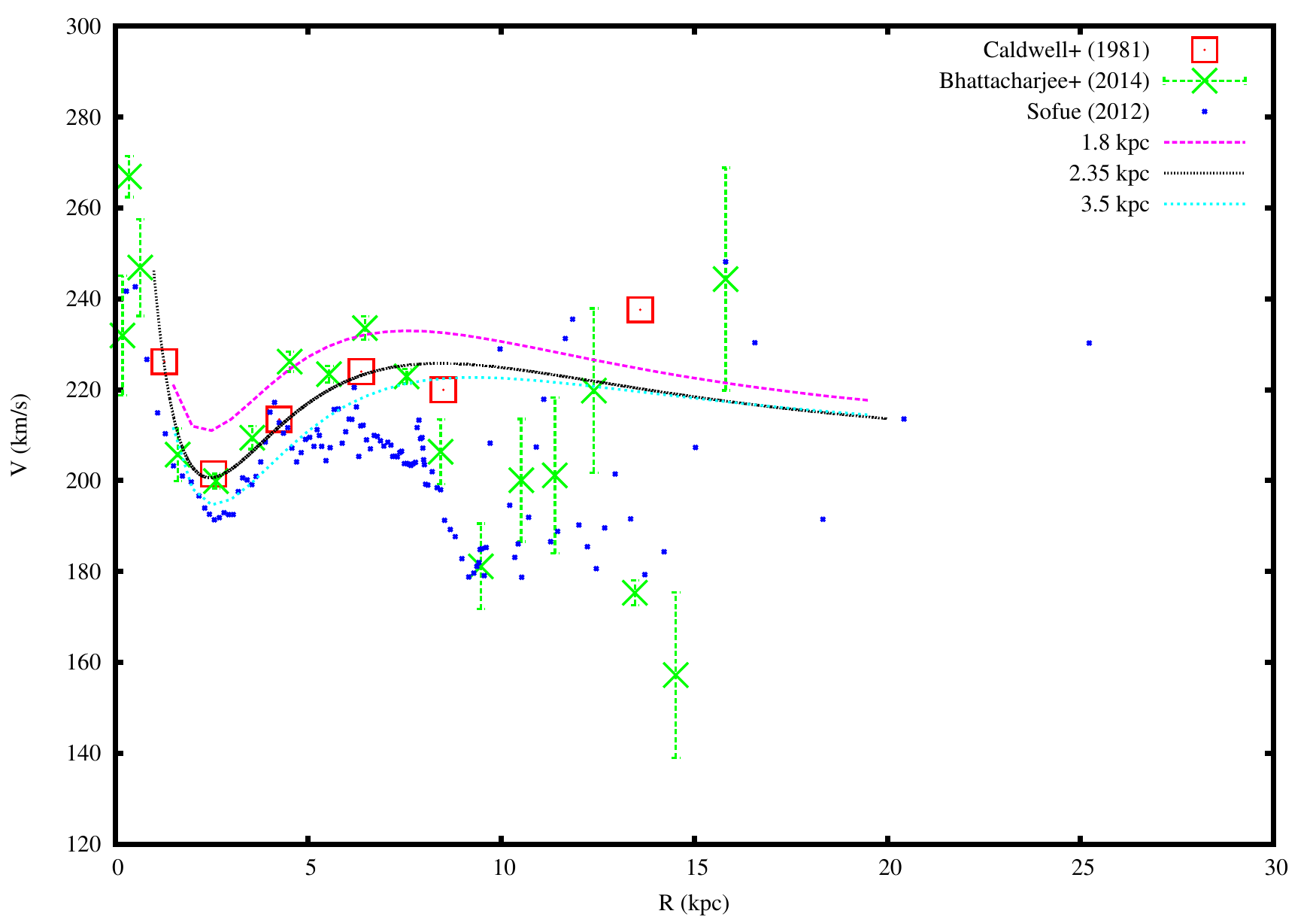}
\caption{Rotation curve of the Galaxy. Observations from Caldwell \& Ostriker (1981) (open squares), Bhattacharjee et al (2014) for R$_o$=8 kpc (crosses), Sofue (2012) blue dots, and model assuming various scale lengths for the thick disc : 2.35 kpc (black line), 1.8 kpc (magenta short dashed line), 3.5 kpc (blue dotted line).  
}
\label{vrot}
\end{figure}

 The rotation curve of the Galaxy is very sensitive to the scale length of the various populations. Using the method reported by \cite{Bienayme1987} we computed the rotation curve for our new potential assuming a thick-disc scale length of 1.8 kpc,  2.35 kpc (from this study) and 3.5 kpc, shown in Fig.~\ref{vrot}, together with observations from the compilation of \cite{1981ApJ...251...61C}, {  from \cite{Bhattacharjee2014}, and from \cite{Sofue2009}. We adopted the thin-disc scale length of 2.17 kpc as determined by \cite{Robin2012a} in the inner Galaxy}.The dark halo density and core radius were fitted on the observed rotation curve following the method described in \cite{Bienayme1987}. A too short scale length for the thick disc (1.8 kpc) implies an over-rotation in the inner galaxy, the disc becomes an over-maximum disc, while the 2.35 kpc thick-disc scale length gives a rotation curve that agrees well with the observed rotation curve.

 \section{Discussion}
\label{Discussion}

In the past 30 years, much controversy emerged after the discovery of the thick disc: the characterisation of the structural parameters of this supposedly minor population was difficult because of the mixture of populations either with the thin disc at low latitudes, or with the halo at high latitudes. The thick-disc scale height determination has  long been found degenerated with the determination of its local density \citep{Reyle2001, Juric2008, DeJong2010}. This caused varying values of the scale height of the literature between ~500 pc to 1200 pc, while the local density was found to be between 1\% to 15\% of the thin-disc local density. Moreover, the scale length was hardly accurately determined, because most studies avoided too low latitudes, which are contaminated by the thin disc and by extinction. 

We now compare our result with recent studies and discuss them with regard to proposed scenarios of thick-disc formation.

\subsection{Thick disc shape, scale height and scale length}

\cite{Reyle2001} have investigated the thick disc and halo shapes using photometry in a limited number of fields available at that time. They found a thick-disc scale height of about 800 pc and a scale length of 2.5 kpc. They reported the strong degeneracy between the scale height and the local density of the thick disc. Considering this degeneracy and returning to their result (Fig. 4) we see that our new solution agrees well with their solution ellipse. The new result is more constrained, the size of the ellipse is now much reduced, and the degeneracy nearly resolved thanks to the wide range of longitudes and latitudes used.

\cite{Du2006} analysed the shape of the thick disc and stellar halo from 21 of the Beijing-Arizona-Taiwan-Connecticut (BATC) Survey  fields. They concluded that there is a thick-disc scale height between 600 and 1000 pc, which reflects the degeneracy between the parameters.

\cite{Juric2008} proposed a new galaxy model with thin disc, thick disc, and halo parameters fitted to SDSS photometric survey. They found a thick disc scale height varying between 600 and 900 pc, depending on the colour bin they considered. They also found the strong degeneracy between the scale height and its local density. However, for the scale length they found a value of 3.6 kpc, longer than ours, but with a considerable range of uncertainty between 2.3 and 6 kpc, and concluded that this parameter is poorly constrained. { They also argued that the thick disc follows an exponential vertically, while we reliably found that the shape is more a hyperbolic secant squared. From SDSS data alone it is difficult to identify that the density law deviates from an exponential at short distances, because of the latitudes of the survey and because most stars close to the Galactic plane are too bright. We combined SDSS with 2MASS in our analysis to identify this feature. Moreover a deviation from the exponential at short distances allowed us to have a more physical law, derivable in the plane, which is useful in dealing with dynamical computation; the exponential shows a singular point in the derivative at z=0.}

\cite{DeJong2010} attempted to fit the SEGUE imaging survey with a smooth model with a thick disc and a stellar halo. They found a thick disc with a scale height of 750$\pm$70 pc and a scale length of 4.1 $\pm$0.4 kpc. The degeneracy between the scale height and the local density is again very strong. The level of accuracy of the fit is 25\% in the mean. Their scale height is compatible with ours, given the uncertainties and degeneracies. However, their scale length is much longer than ours and incompatible at the 3 sigma level. Because the range of longitude and latitude studied is much larger in our case, we expect that we better trace the thick disc population in the inner Galaxy. Moreover, they did not considere a flaring thick disc, which can likewise be reflected in the estimated scale length.

\cite{Chang2011} analysed the 2MASS survey to determine the Galactic structure parameters for the thin disc, thick disc, and halo. They avoided the low latitudes $|b|<30$\deg and again found the high degeneracy between density and scale height of the thick disc, and between the halo axis ratio and its power-law exponent. Their result is compatible with ours, but much more degenerate.
% j'en suis lˆ...
\cite{Bovy2012a} studied SEGUE G-dwarf spectroscopic samples using a mono-abundance decomposition. They determined the selection function of the survey, distances, and estimated the scale length and scale height of sub-populations with different iron and alpha-element abundances. They deduced that the thick disc has a scale height of about 655 pc and a scale length of 1.96 kpc. They claimed that an exponential law can fit their data, but they did not try another shape. Their study covered a limited Galactocentric distance range, only to 3 kpc from the Sun position. In their data the scale heights seem to be continuous when one varies the alpha and iron abundances. However, it has been seen since their study that there is a clear discontinuity in $\alpha$ element abundances between the thin disc and thick disc \citep{Adibekyan2013, Haywood2013, Bergemann2014, Recio-Blanco2014}. 

 Despite the differences in methods, our result agrees well with the determination of \cite{Bovy2012a}  of the thick-disc scale length. Their scale length of the $\alpha$-old sample is found to be ~2 kpc. Here we find a mean scale length of 2.3 kpc, but more specifically, 2.0 kpc for the younger thick disc, and 2.6 kpc for the older part. Our values of the shape parameters are most likely better constrained than those of \cite{Bovy2012a} because they used a sample that reached only 2-3 kpc, while our sample reached 10 kpc distances for the thick disc. In particular, they were unable to constrain the flare. 
 
 { This is different for the thin-disc population, which they found to have a longer scale length of 3.5 kpc. However, as can be seen in their Fig. 5, the uncertainty on the thin-disc scale length
 is large, probably because of scarcity of the G dwarf sample and the small distance range covered (about 2 kpc for the selection at $|z|<250$ pc). We are investigating the radial thin-disc scale length using 2MASS data in the Galactic plane  (Amores et al, in prep). Preliminary results show that the thin-disc scale length varies with age and that its value is robustly constrained to be 2.2 kpc for stars older than 5 Gyr in the thin disc}. 
 
We are also consistent with \cite{Bensby2011}, who evaluated the scale length to be 2 kpc, but with samples very limited in size and without estimating of the error bar, although it is not sure that the two methods trace exactly the same population. Our thick disc has a very similar short scale length to that of the old thin disc, as is also observed in most external galaxies \citep{Yoachim2006}.
Hence we conclude that the old thin disc and young thick disc have similar short scale lengths, but their scale heights differ by about a factor of 2.

Most recently, \cite{Bovy2013} estimated on the ground of the rotation curve that the total scale length of the disc is 2.15 kpc, which agrees very well with our two populations of thin and thick disc. This value of 2.15 kpc would be difficult to  justify with a thin-disc scale length of 3.5 kpc and a thick disc of less than 10\% of the thin disc with a scale length of 2 kpc.

\subsection{Thick-disc flare}

{ Our analysis shows that the thick disc is flaring in its outskirts, but only for the oldest (and less massive) component. The main (younger) thick disc is found to be much flatter. For the old thick disc the flare parameters also depends on the halo shape. Therefore this is probably related to the collapse phase between the halo and old thick disc.

\subsection{Thick disc in the inner galaxy}

While a detailed analysis of the inner region is not the purpose of this article, we stress that the contribution of the thick disc to the populations in the bulge region explains the stellar components identified in the ARGOS survey \citep{Ness2012,Ness2013a,Ness2013b} very well. These data show that different populations overlap in the inner region of the Milky Way, and at moderate latitudes their component C has metallicity, alpha abundances, and velocities very similar to the old thick disc we found, while component B could be a counterpart of our younger thick disc. Most recently, \cite{DiMatteo2014} analysed simulations of the bulge evolution and showed that components B and C identified by ARGOS survey can also be identified with the young and old thick discs. A more detailed analysis of the metallicity and velocities distributions in the inner region with newly available spectroscopic surveys will be performed in the near future using the new population model proposed here. 
}

\subsection{Age and metallicity}

Our estimate of age and metallicity only relies on photometric observations. This is mainly the shape of the luminosity function and the $g-r$ colour of the turnoff determined in our study, not directly the age and metallicity. Spectroscopic surveys are more efficient in determining the metallicities, and the age can be obtained from Bayesian analysis when the spectroscopy is accurate enough \citep{Haywood2013}. 

Even though they are less accurate, the thick disc and halo metallicities found in our study agree well with the recent determination from the BATC survey \citep{Peng} which found [Fe/H] $\approx$ -1.5 in the distance from the Galactic plane $|z| > 5 $kpc, corresponding to the halo and [Fe/H] $\approx$ -0.7 in the region $2 < |z| < 5 $kpc, corresponding to the thick disc. They did not find evidence of a vertical gradient for these populations.

From the low-latitude fields of the SEGUE survey, \cite{Cheng2012} estimated the mean thick-disc metallicity of turnoff thick-disc stars to be [Fe/H] $\approx$ -0.5 dex and found no radial gradient within this population. \cite{Schlesinger2012} analysed samples of G and K dwarfs in higher latitude SEGUE fields. They found a mean metallicity peaking at about -0.5 dex for stars at distances larger than 1.5 kpc from the plane. At this distance it is still possible that the sample is contaminated by the thin disc, which would enlarge the distribution towards higher metallicities.

In our analysis we did not assume any radial or vertical metallicity gradients. This simple assumption may be reconsidered in the future, but broad-band photometry alone is not accurate enough to determine a subtle change in the metallicity. 

Concerning ages, our analysis is as reliable as are the isochrones used. However, we clearly improve the fit (in particular the distribution at the turnoff in CMDs) by having a longer star formation period for the thick disc. Our thick disc can have formed stars approximately from 12 to 10 Gyr ago, but the star formation efficiency was lower at the beginning, the old episode forming only 13\% of the thick-disc stars in our best model.
\cite{Haywood2013} determined the age of the thick disc from the local sample of \cite{Adibekyan2013} and deduced that the thick-disc age ranges between 9 and 12 kpc, which agrees with our finding, although our modelling is too coarse to be able to distinguish a continuous formation from two distinct events. We plan to use the new model scheme described in \cite{Czekaj2014} to study whether a continuous formation period of the thick disc can represent the data better.

\subsection{Thick-disc formation scenario}
{ 
The thick-disc formation scenario can be investigated in the light of the different observational characteristics. Judicious elements to constrain the formation scenario are the mass distribution, the star formation efficiency, the chemical abundances, and the kinematics. We did not consider this last aspect in this paper, postponing it to a future paper. In recent years, several spectroscopic surveys have given new clues for the thick-disc formation scenario by giving access to the $\alpha$ abundances, which has been shown to be a more efficient parameter for distinguishing the thin disc from the thick disc. However, spectroscopic samples are scarce and suffer from severe selection effects. This is why we used photometric data, which allowed us to explore much larger distances, with very significant stellar densities and only small (and easy to correct) selection biases. We did not make any selection on the sample based on metallicity and abundances, or on kinematics. Hence, we limited the bias in the selection to the magnitude-selected sample, which is very easy to simulate by population synthesis. We showed that using this approach we were able to reproduce the star counts very well and quantitatively reproduced the distributions in colour-magnitude diagrams on a wide range of longitudes and latitudes. We can therefore use this new characterisation of the thick disc to constrain its formation scenario.

It has been claimed that the thick disc cannot be considered as a separate population from the thin disc \citep{Bovy2012a}.
Radial migration was also proposed to explain the formation of thick disc strictly from migration in the thin disc \citep{Schonrich2009a,Schonrich2009b}. But \cite{Minchev2014} showed that radial migration is not efficient enough to form a thick disc with the observed characteristics. Moreover, the $\alpha$ sequences in the data used by \cite{Schonrich2009a} did not show a clear separation of the low-$\alpha$ (thin disc) from the high-$\alpha$ (thick disc) which now is well established.

The local sample of FGK stars from \cite{Adibekyan2013} has shown a clear separation on thin and thick disc sequences in the 
$[Fe/H]$ versus $[\alpha/Fe]$ plane. This separation is also well established in the APOGEE sample, while the lower resolution SEGUE survey did not show it. These new high-resolution data are able to separate the thin disc from the thick disc and also allow better determination of distances and ages. \cite{Haywood2013} evaluated the age, iron and alpha abundance, and velocities of thin and thick disc and proposed a scenario for the formation of the thick disc. They claimed that the observations are not in favour of an inside-out disc formation, nor of a significant churning source of radial migration. Using the assumption \citep[on the basis of ][]{Bovy2012a}  that the thick disc has a shorter scale length than the thin disc, they interpreted this by a formation of the thick disc in a smaller inner galaxy at early epoch, with a scale height that diminishes with time and during 4 to 5 Gyr. In this scenario the inner thin disc forms later on the ground of the metallicities left by the thick-disc enrichment [Fe/H] ~0 $\pm$ 0.1 dex  and [$\alpha$/H] ~0-0.1 dex, while the outer disc was born earlier during the thick-disc formation. In their scenario we would expect an age dependence of the thick disc on Galactocentric radius, which we do not observe, within the accuracy of our photometric study. 
Several attempts have been made to propose that thick discs are formed by
 minor mergers. Several simulations of minor mergers on pre-existing discs have been performed, for instance by \cite{Quinn1986,Villalobos2008}. In this last paper the authors showed how the initial disc is heated and described the characteristics of the resulting thick disc, which depended on the parameters of the merging satellite. In particular, they demonstrated that the scale length of the final thick disc is always slightly more extended than the original disc. A significant flare is also observed in these thickened discs. These two characteristics are not observed in this study. 

An alternative scenario has been proposed by \cite{Bournaud2009}, who showed that thick discs can be efficiently formed from gas-rich turbulent giant clumps at high z. They explored the different characteristics of thick discs formed by mergers or formed by this new process. They demonstrated that thick discs formed by mergers have significant flares (in agreement with the merger simulations), but thick discs formed by giant clumps do not flare. They obtained a constant scale height at all radii. Another interesting feature is that the scale lengths of the thick disc is comparable with that of the thin disc, as was found in our study. Thus the characteristics of the thick disc in the Milky Way is well described by this scenario of formation from turbulent giant clumps at about z $\approx$ 2.

\cite{Martig} have followed the evolution of a set of simulations similar to the Milky Way, some of them with mergers, some without. They showed that when the merger history is significant, the oldest disc populations are significantly flared. Their simulation, called g92, is very flat (i.e. has a constant scale height) to at least 20 kpc. Some slight flaring is present in the remote outskirts, probably because of outwards migration of stars, as in the simulations of \cite{Minchev2012}. The simulations with significant merging history show a stronger flare in the outer disc, which is also stronger with time.
It has been noted that tidal effects of small satellites could alter the outer parts of discs \citep{Bird2013}, another process that might explain several structures observed in the anticentre region of the Milky Way, such as the Monoceros ring or the Canis Major overdensity.

The simulations of \cite{Bird2013} explain well an anti-correlation between scale length and scale height, as claimed by \cite{Bovy2012b}, by an inside-out formation of the disc. In contrast, \cite{Stinson2013}  found a correlation between the scales, as we see in our study. In the simulations of \cite{Martig} both cases can occur, depending on the individual simulation considered. In agreement with our result, the thick disc itself does not show an anti-correlation, but rather a correlation of its scales. 

Thus, our characterisation of the thick-disc shape and star formation history favours a formation of the thick disc following the scenario of \cite{Bournaud2009}, where the stars are formed during a period at high redshift where the gas is turbulent enough to resist to the gravitational collapse of the gas to a thin disc. During this period, the thick disc might be slightly contracting, explaining why the scale height and scale length are both decreasing with time. The flare that is not present in the main thick disc is well in agreement with this scenario, while other scenarios such as radial migration or thin-disc heating by mergers would have produced a significant flare in the outer thick disc \citep{Minchev2012,Roskar2013}. Moreover, it should be noted that in external galaxies, most thick discs do not show outer flares \citep{Dalcanton2002}, also favouring the scenario of \cite{Bournaud2009}.

In this scenario the thick-disc population is well mixed such that no radial metallicity gradient should be present in the high $\alpha$ sequences, as is observed in recent spectroscopic surveys \citep{Cheng2012,Anders2013,Hayden2014, Recio-Blanco2014}. After the formation episode of the thick disc, slightly outside-in, but mainly sustained by turbulence, the thin disc itself can form inside-out from a standard process including radial migration. This could explain both the correlation of scales during the thick-disc episode, and their anti-correlation during the thin-disc formation. 

}
\subsection{Halo shape}

\cite{Sesar2011} analysed CFHT-LS survey data to study the structure of the stellar halo. They found that the halo follows a power law with an axis ratio of 0.70, agreeing well with our analysis. However, they found that the exponent varies between 2.6 and 3.8 depending on the radius (with a transition at about 28 kpc). Their analysis was limited to four directions, thus is less constraining than ours. But most probably the difference in the power law exponent we found (3.3 instead of 2.6) is due to the fact that among the four regions in the CFHT-LS, at least two are contaminated by halo streams. This was the reason why we did not use these fields in our analysis.

With the hypothesis that the halo follows a Hernquist shape, we estimate the axis ratios to be  q=0.77, but the exponent is smaller (2.76) than in the case of a power-law density (n=3.3). The different formulae explain this difference. We attempted to fit a triaxial Hernquist halo, but the parameters (angles and intermediate axis ratio) are ill-determined. Therefore we did not find any clear evidence of triaxiality. 
\cite{Newberg2007} used SDSS photometric data to fit a Hernquist halo. Assuming a centered halo, as we did, they found an axis ratio p=0.73 and q=0.60, with an angle $\theta$=70\deg. They did not give a confidence interval for their values. However, they used the entire SDSS northern survey (DR6 and DR7) to derive these parameters. Here we considered a restricted portion of the survey to avoid contamination by the main streams, while results of \cite{Newberg2007} can be influenced by these streams, which could give an impression of a significantly triaxial halo even if the smooth part of the halo is not.

From an analysis of the SEGUE survey, \cite{Carollo2008} claimed that the halo is dual, with the inner halo being more metal rich, having a prograde motion and flattened spatial resolution, while the outer halo would be more metal poor, having retrograde motion and being more spherical. \cite{Kafle2013} also argued for a dual halo from an analysis of BHB stars, where they considered the kinematics for different metallicity groups. However, this is still a matter of controversy, as \cite{Fermani2013} and \cite{Schonrich2014} claimed that this interpretation of SEGUE data is biased and that there is no clear evidence yet for a dual halo. Here we tried to determine whether the data are able to indicate a change of slope and/or a change of flattening at some Galactocentric distance.  We assumed both parts are power laws, but with different exponents and axis ratios. But after many iterations on each MCMC run the values of the distance of the break between two halos and also the axis ratio and exponent of the outer halo did not converged. 
The distances histogram of the subgiants in the sample shows that 95\% of the simulated stars have a Galactocentric distance within 20 kpc  and a z distance within 25 kpc, in the SDSS sample.
The number of distant stars ($>$ 30 kpc) is too small in our data set to allow us to determine the shape of the outer halo accurately enough.

\section{Conclusion and perspectives}

{Our study combined the two major digital surveys conducted in recent years, 2MASS in the infrared and SDSS in the visible, to characterise the distribution of stars formed during the early phases of Galaxy formation. Using the stellar population approach, we reconsidered the thick disc and halo structures and adjusted their parameters efficiently. We conclude with a set of reliable conclusions, which are as follows:

\begin{itemize}
\item {Considering a single burst of formation for the thick disc, its density law is characterised by a shape that more resembles a hyperbolic secant squared than an exponential. The whole shape can be characterised by a parabola to a distance of about 660 pc from the plane followed by an exponential  scale height of 535 $\pm$ 50 pc, or alternatively by a $sech^2$ shape with a scale height of 470 pc. The scale length is  found to be 2.3$\pm$ 0.1 kpc and it is found to be flaring in the outer Galaxy. It is about 7.9\% of the thin disc locally. This value may depend on the IMF and on the binary fractions, if they differ in the two populations.
\item Considering a longer time of formation, we obtain a better fit if the thick-disc formation period extends between 12 and 10 Gyr.  
We find that the older thick disc has a significantly larger scale height and scale length than the younger thick disc, the sech$^2$ scale height reaches 800 pc at 12 Gyr down to 350 pc (twice that of the thin disc) at 10 Gyr, and the exponential scale length from 3 kpc to 2 kpc for the same period.   The degeneracy between thick-disc scale height and local density found in many studies can be explained both by the selection of a limited range of longitudes and latitudes, and by the fact that during the thick-disc formation the time laps of the collapse produced a correlation between scale length and scale height, which reflects the contraction. 
\item The inner part of the thick disc significantly contributes to the stellar populations in the inner Galaxy and was misinterpreted as a thick bulge or a flattened classical bulge in \cite{Robin2012a}. The thick-disc contribution in bulge fields can explain population C described in the analysis of the metallicity distribution in the Argos survey by \cite{Ness2013a}.
\item No flare is found during the major episode of thick-disc formation.}
\end{itemize}

These results favour a formation scenario proposed by \cite{Bournaud2009} where thick-disc stars are formed during a period at high redshift where the gas is turbulent enough to resist to the gravitational collapse of the gas to a thin disc. During this period, the thick disc might be slightly contracting, which explains why the scale height and scale length both decrease with time. The flare, which is absent in the main thick disc, agrees well with this scenario, while other scenarios such as radial migration or thin-disc heating by mergers would have produced a significant flare in the outer thick disc \citep{Minchev2012,Roskar2013}. Noteworthy in external galaxies, most thick discs do not show outer flares \citep{Dalcanton2002}, which also favours the scenario of \cite{Bournaud2009}.
}

Our analysis of the shape and mass of the thick disc needs to be combined with spectroscopic data to improve the determination of age and metallicities of the thick-disc population. Kinematical data will also help to understand its formation.
In a future paper we will present kinematical constraints from SDSS radial velocity and proper motions, and analyse SEGUE and APOGEE data in terms of metallicity gradients. We also propose to use the new model from \cite{Czekaj2014}, that has so far only been applied to the thin-disc population, to improve the modelling of the thick disc, and to compare a more detailed star formation history with various scenarios of formation and evolution.

\begin{acknowledgements}
We acknowledge the support of the French Agence Nationale de la
Recherche under contract ANR-2010-BLAN-0508-01OTP. BGM simulations
were executed on computers from the Utinam Institute of the
Universit\'e de Franche-Comt\'e, supported by the R\'egion de
Franche-Comt\'e and Institut des Sciences de l'Univers (INSU).

JF acknowledges support by he Spanish Ministerio de
Econom\'ia y Competitividad (MINECO; grant AYA2010-21322-C03-02).

    Funding for the SDSS and SDSS-II has been provided by the Alfred P. Sloan Foundation, the Participating Institutions, the National Science Foundation, the U.S. Department of Energy, the National Aeronautics and Space Administration, the Japanese Monbukagakusho, the Max Planck Society, and the Higher Education Funding Council for England. The SDSS Web Site is \url{http://www.sdss.org/}.

    The SDSS is managed by the Astrophysical Research Consortium for the Participating Institutions. The Participating Institutions are the American Museum of Natural History, Astrophysical Institute Potsdam, University of Basel, University of Cambridge, Case Western Reserve University, University of Chicago, Drexel University, Fermilab, the Institute for Advanced Study, the Japan Participation Group, Johns Hopkins University, the Joint Institute for Nuclear Astrophysics, the Kavli Institute for Particle Astrophysics and Cosmology, the Korean Scientist Group, the Chinese Academy of Sciences (LAMOST), Los Alamos National Laboratory, the Max-Planck-Institute for Astronomy (MPIA), the Max-Planck-Institute for Astrophysics (MPA), New Mexico State University, Ohio State University, University of Pittsburgh, University of Portsmouth, Princeton University, the United States Naval Observatory, and the University of Washington.

This publication makes use of data products from the Two Micron All Sky Survey, which is a joint project of the University of Massachusetts and the Infrared Processing and Analysis Center/California Institute of Technology, funded by the National Aeronautics and Space Administration and the National Science Foundation.

This research has made use of the VizieR catalogue access tool, and CDSclient, CDS, Strasbourg, France.

\end{acknowledgements}

\bibliographystyle{aa} % style aa.bst
\bibliography{thick} % YOUR REFERENCES WD.bib

\begin{appendix}
\section{Comparisons between the best fit model and photometric data}

We present here complementary figures that shows the comparison between our best-fit model with data from the SDSS and 2MASS surveys as histograms in different fields and different magnitude ranges.

Fig.s A.1 to A.5 show the comparison between the model and SDSS data in a variety of longitudes and latitudes.

\begin{figure*}
\includegraphics[width=14cm,angle=0]{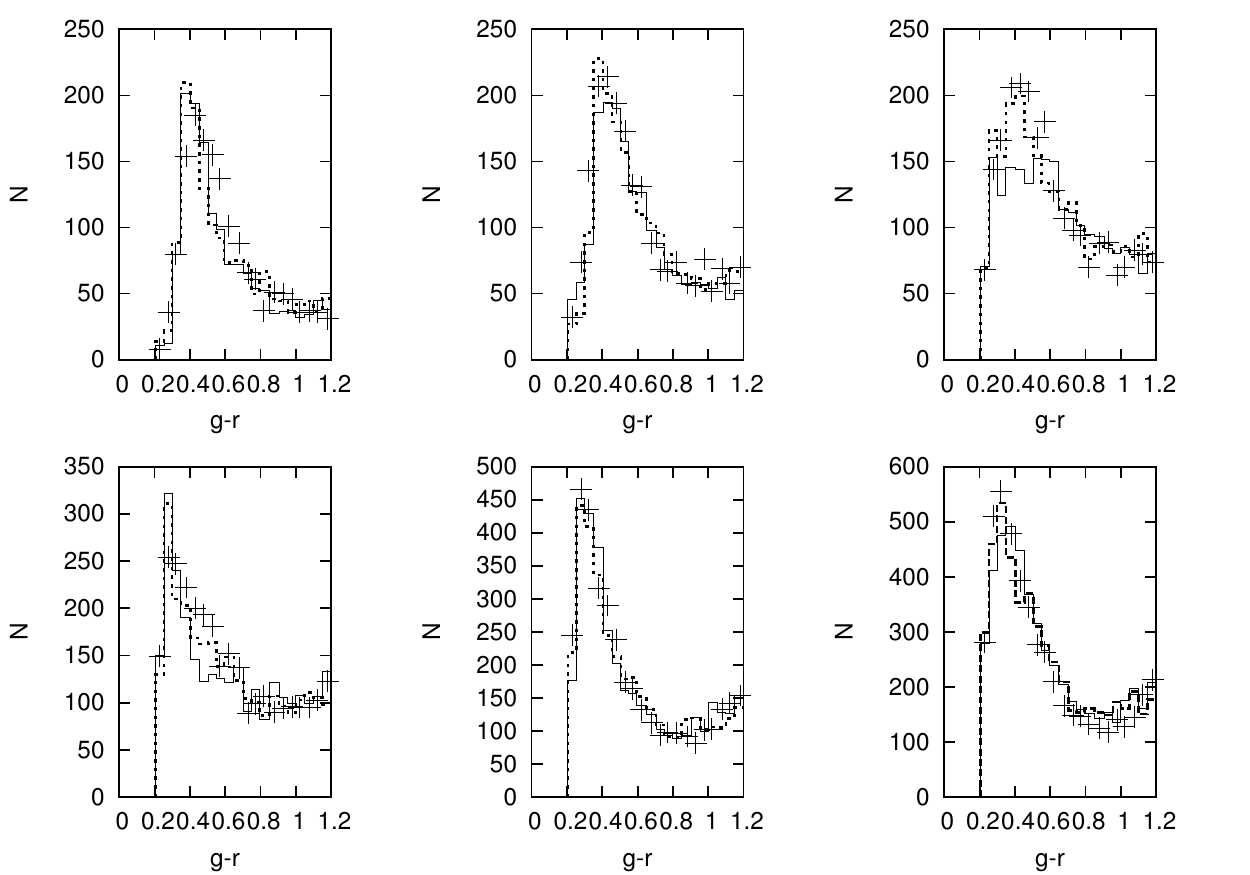}

\caption{Comparison of the best-fit thick-disc model with star counts from SDSS data in the field at longitude 100\degr, latitude 75\degr. Each panel represents a different magnitude range, from 15 to 21 in r from top left to bottom right. Data are plotted as plus signs, the model with single-formation episode in the thick disc as solid lines, the model with two formation episodes as dotted lines.}
\label{histo-sdss51}
\end{figure*}

\begin{figure*}
\includegraphics[width=14cm,angle=0]{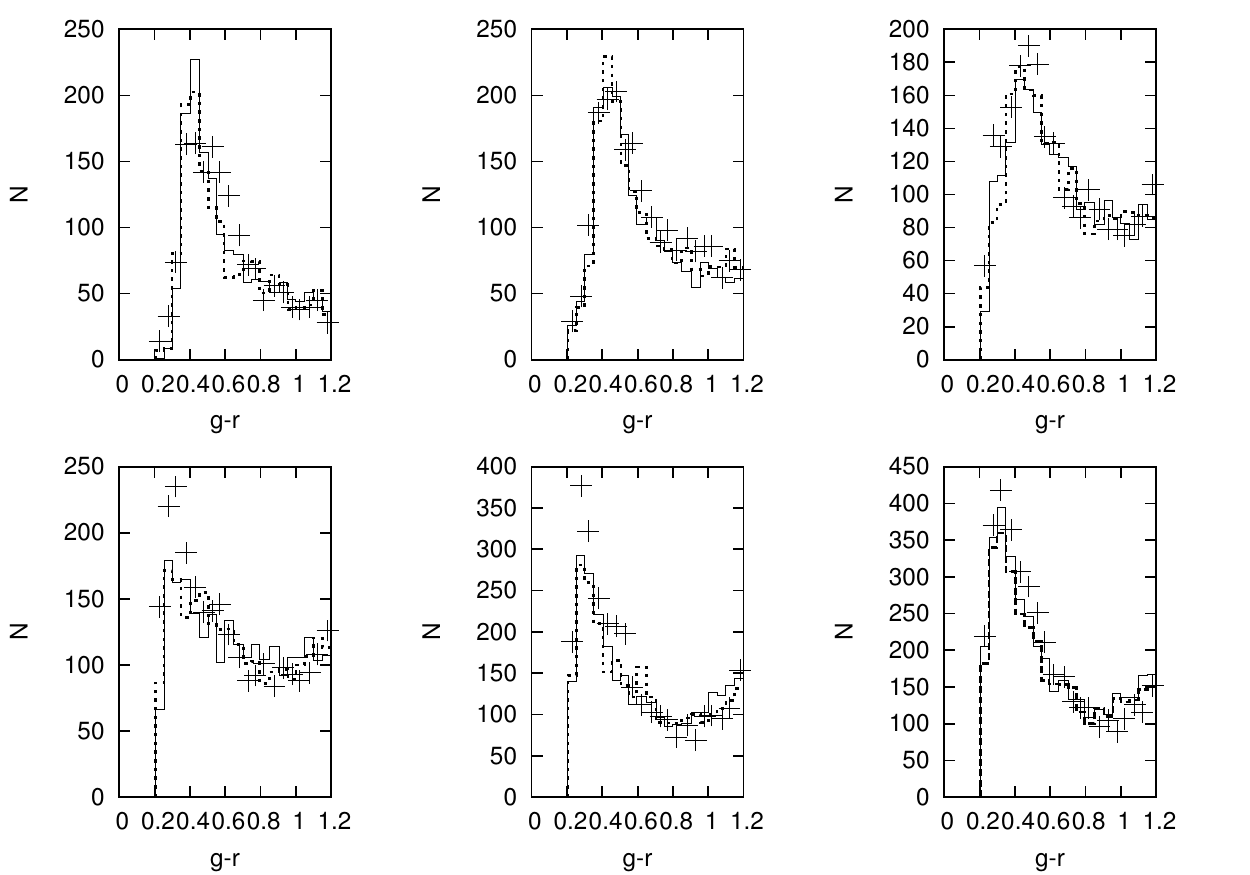}
\caption{Same as Fig. A1 for the SDSS field at longitude 180\degr, latitude 56\degr.}
\label{histo-sdss42}
\end{figure*}

\begin{figure*}
\includegraphics[width=14cm,angle=0]{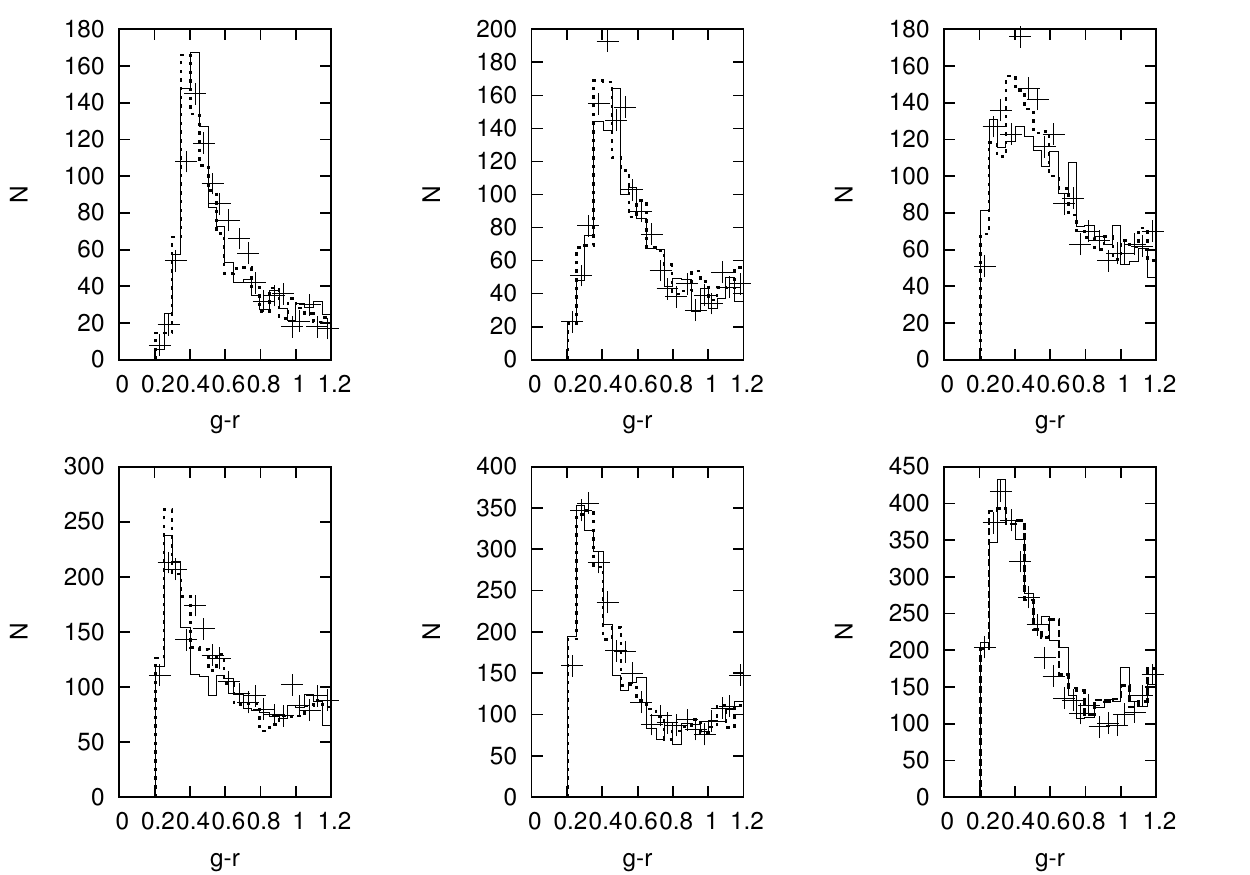}
\caption{Same as Fig. A1 for the SDSS field at longitude 65\degr, latitude 78\degr.}
\label{histo-sdss82}
\end{figure*}

\begin{figure*}
\includegraphics[width=14cm,angle=0]{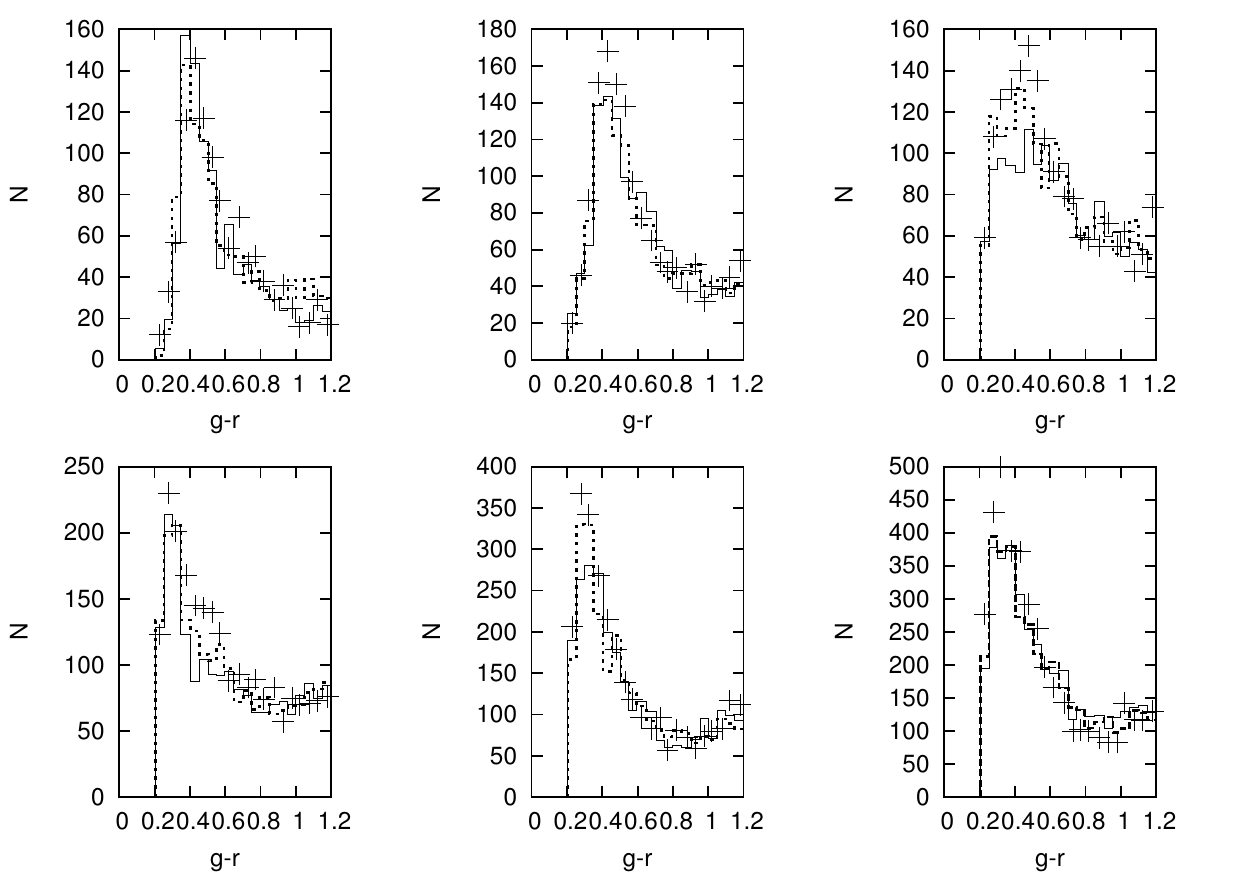}
\caption{Same as Fig. A1 for the SDSS field at longitude 222\degr, latitude 89\degr. }
\label{histo-sdss92}
\end{figure*}

\begin{figure*}
\includegraphics[width=14cm,angle=0]{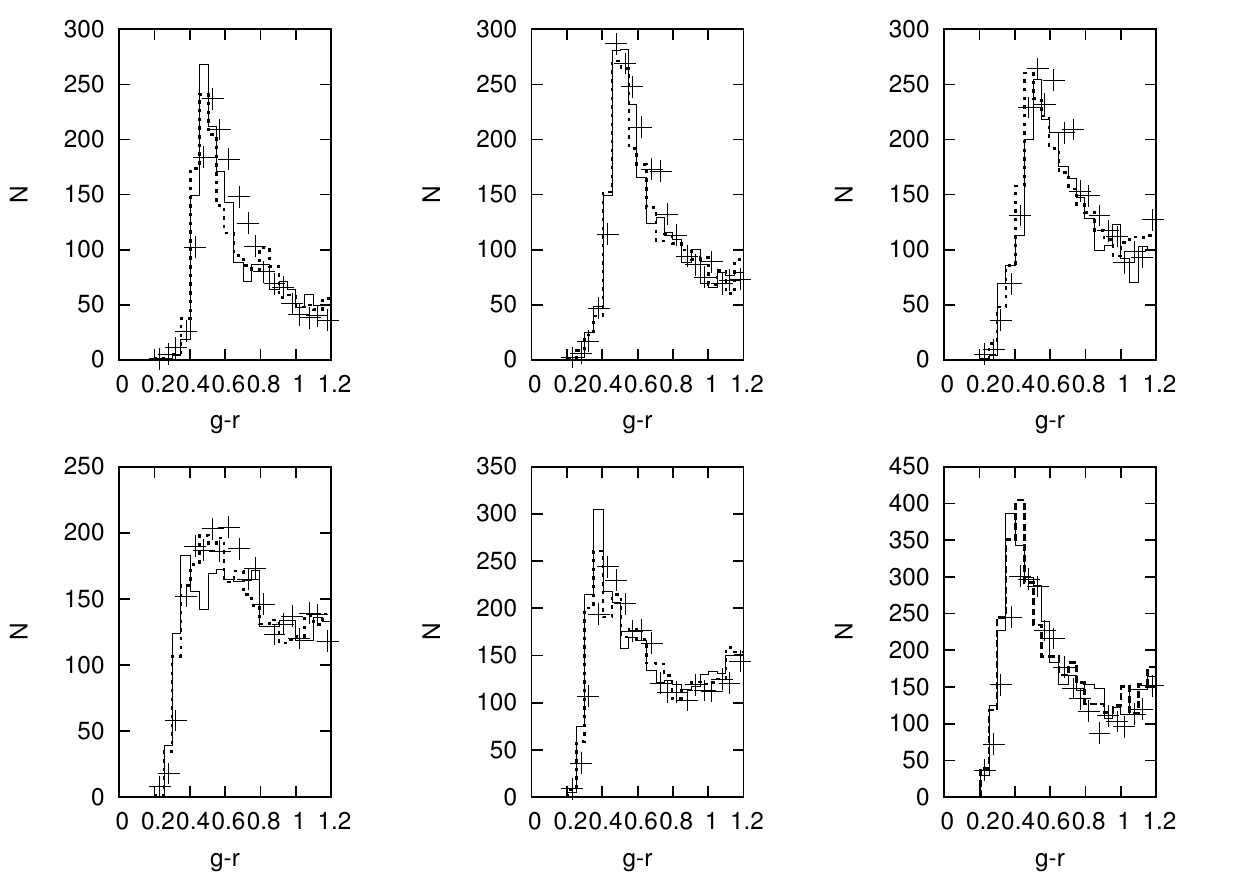}
\caption{Same as Fig. A1 for the SDSS field at longitude 116\degr, latitude -51\degr. }
\label{histo-sdss100}
\end{figure*}

Fig. A.6 to A.9 show the comparison between the model and 2MASS data in a variety of longitudes and latitudes.

\begin{figure*}
\includegraphics[width=14cm,angle=0]{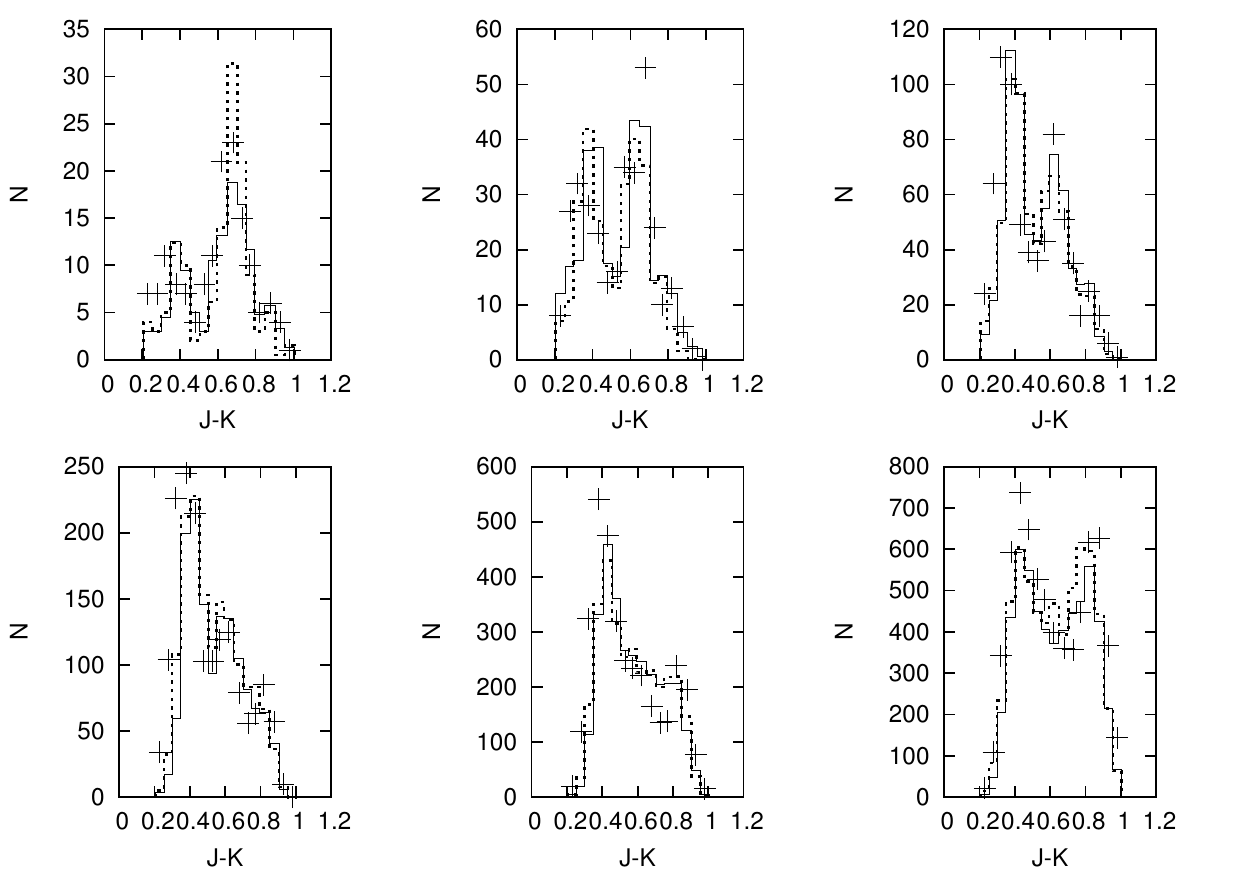}
\caption{Comparison of the best-fit thick-disc model with star counts from 2MASS data in the field at longitude 110\degr, latitude -31\degr. Each panel represents a different magnitude range, from 9 to 14 in K from top left to bottom right. Data are plotted as plus signs, the model with single-formation episode in the thick disc as solid lines, the model with two formation episodes as dotted lines.}
\label{histo-2mass121}
\end{figure*}

\begin{figure*}
\includegraphics[width=14cm,angle=0]{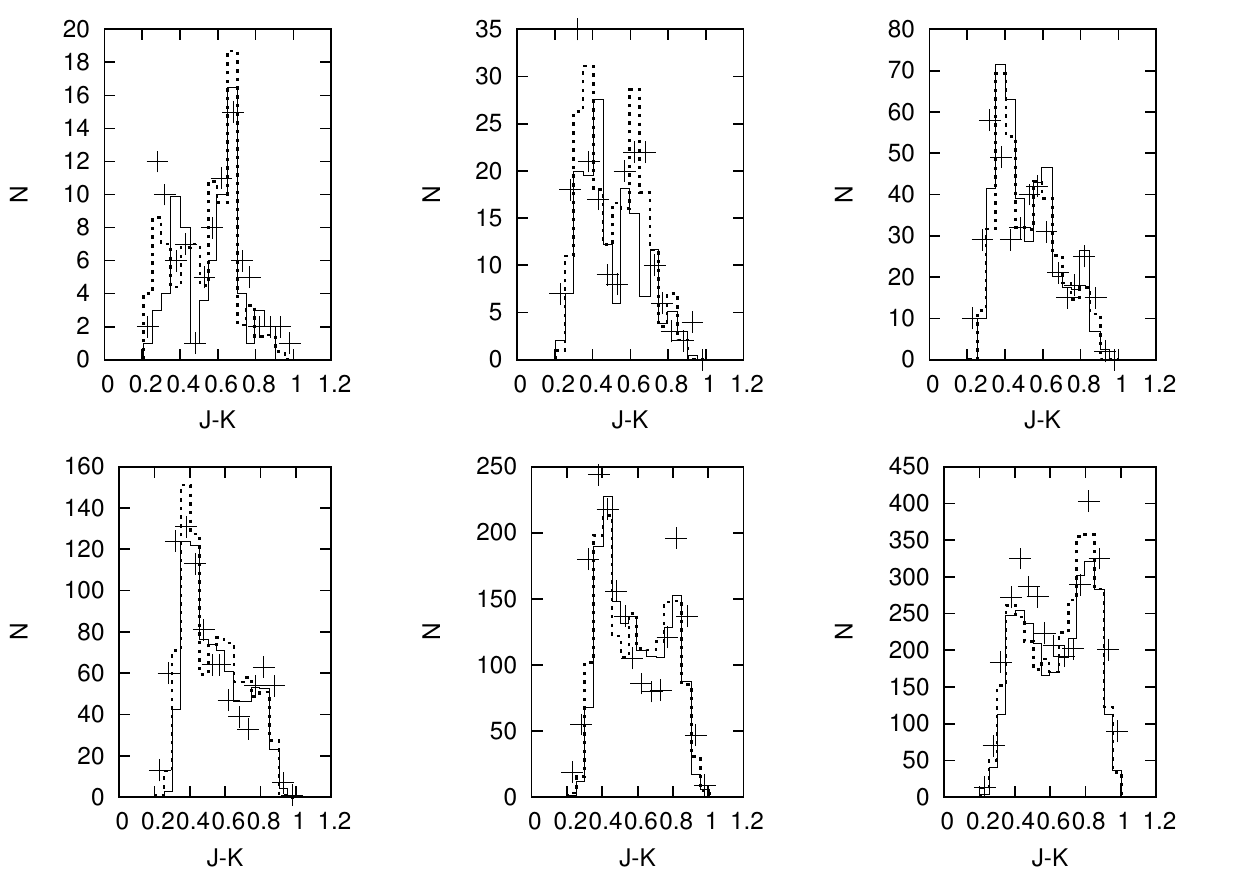}

\caption{Same as Fig. A6 for the 2MASS field at longitude 198\degr, latitude 52\degr. }
\label{histo-2mass125}
\end{figure*}

\begin{figure*}
\includegraphics[width=14cm,angle=0]{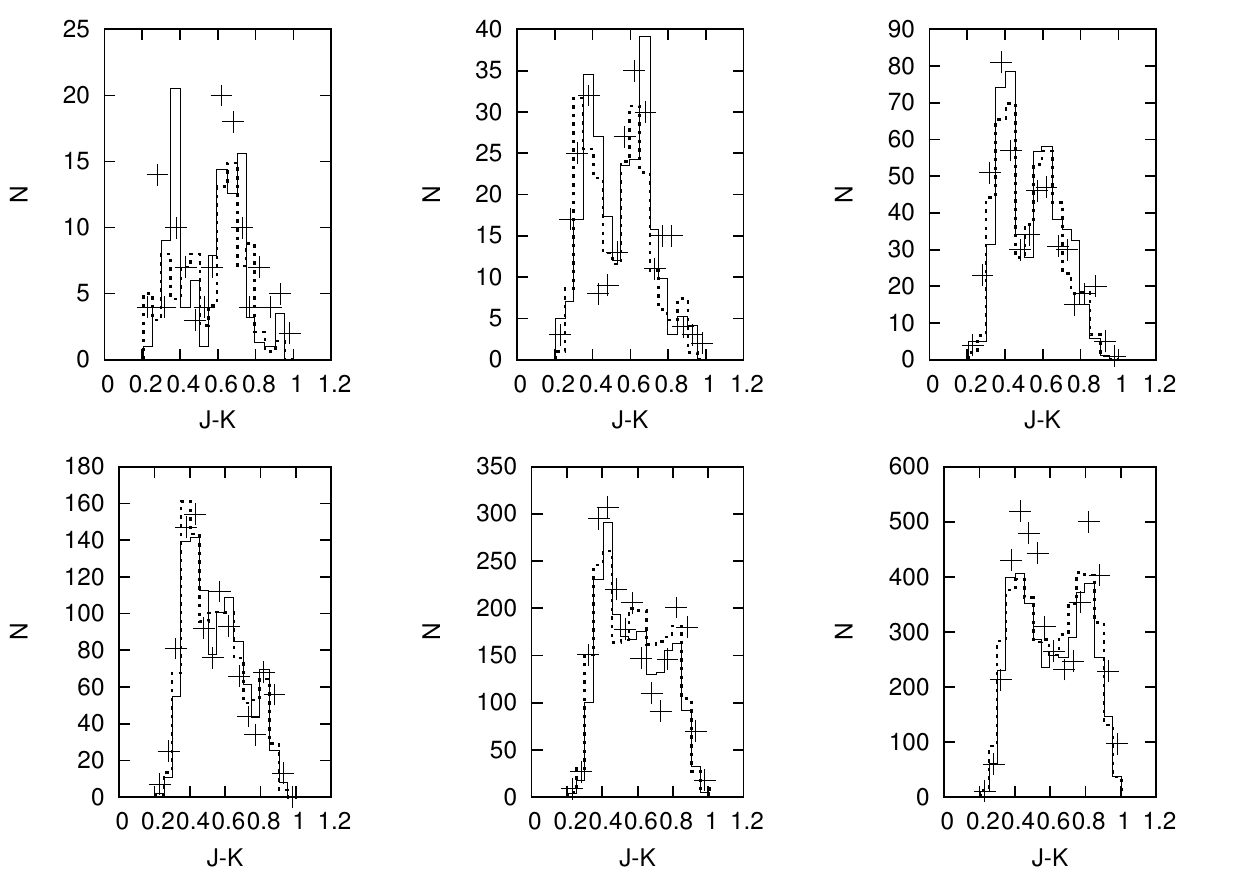}

\caption{Same as Fig. A6 for the 2MASS field at longitude 78\degr, latitude 47\degr.}
\label{histo-2mass137}
\end{figure*}

\begin{figure*}
\includegraphics[width=14cm,angle=0]{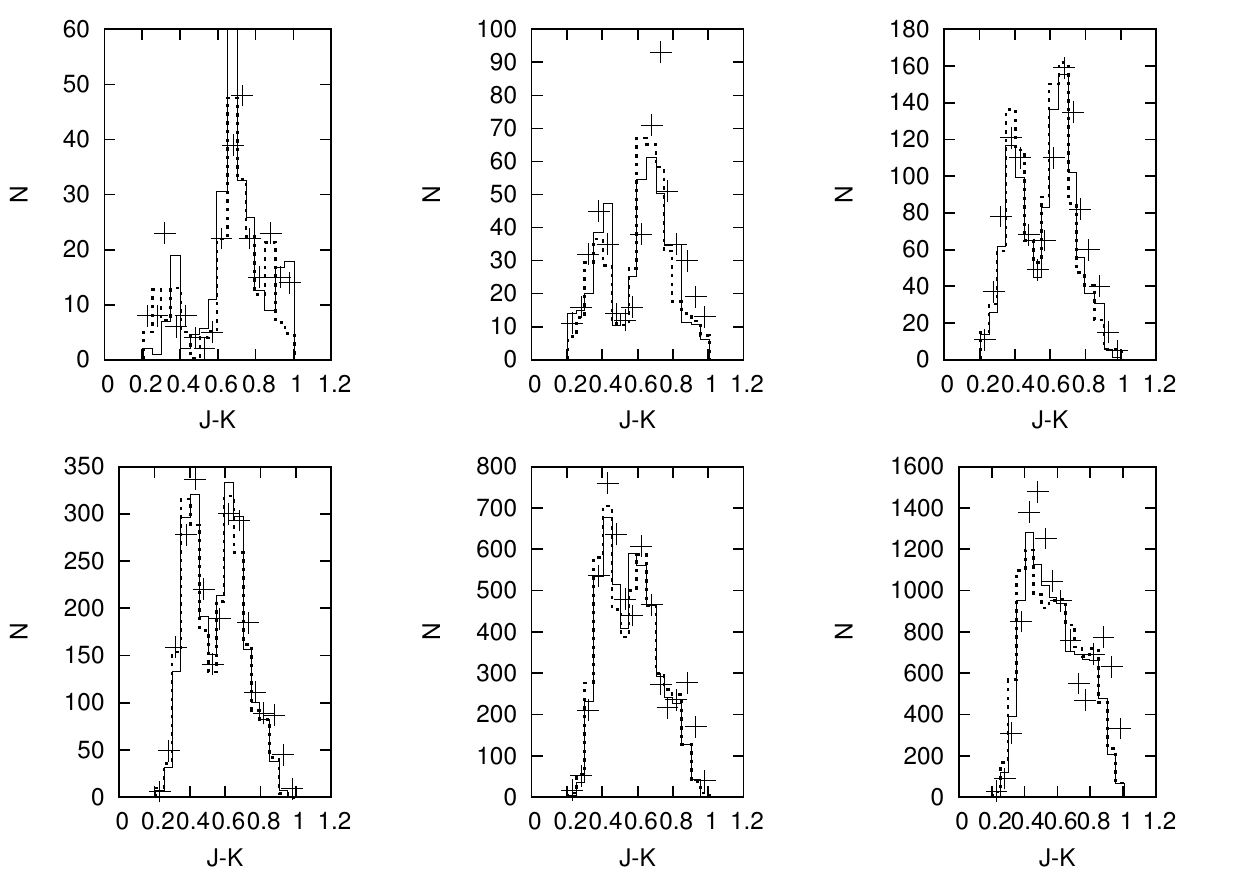}

\caption{Same as Fig. A6 for the 2MASS field at longitude 272\degr, latitude -44\degr. }
\label{histo-2mass152}
\end{figure*}

\end{appendix}

\end{document}